\documentclass[11pt,a4paper]{article}
\pdfoutput=1

\usepackage{epsf,epsfig,subfigure,graphicx,amsmath,amssymb}
\usepackage{amsfonts}
\usepackage{amsmath}
\usepackage{latexsym}
\usepackage{color}
\usepackage{accents}
\usepackage{dhucs}
\usepackage{verbatim}
\usepackage{cite}
\usepackage{xcolor}
\usepackage{slashed}
\usepackage{float}
\usepackage{notes2bib}
\usepackage{hyperref}
\usepackage{wasysym}
\usepackage{bbold}
 
\def\bea{\begin{eqnarray}}
\def\eea{\end{eqnarray}}

\newcommand{\ffdual}{F_{\mu\nu}\widetilde{F}^{\mu\nu}}

\newcommand{\todo}[1]{}

\def\GeV{\,{\rm GeV}}

\def\eV{\,{\rm eV}}

\def\GeV{\,{\rm GeV}}

\def\eV{\,{\rm eV}}

\def\eps{\epsilon}
\def\Rs{R_\star}

  \def\g{\gamma} 

\def\g{\gamma}    
  \def\t{\tau} \def\l{\lambda}

 \newcommand{\Ocal}{{\mathcal O}}

\newcommand*{\affaddr}[1]{#1} 
\newcommand*{\affmark}[1][*]{\textsuperscript{#1}}

\def\Weizmann{\small{Department of Particle Physics and Astrophysics, Weizmann Institute of Science, Rehovot 761001, Israel}}

\def\Mainz{\small{Helmholtz Institute Mainz, Johannes Gutenberg University, Mainz 55099, Germany}}
\def\Berkeley{\small{Department of Physics, University of California, California 94720-7300, USA}}
\def\sw{\small{School of Physics, University of New South Wales, Sydney 2052, Australia}}

\begin{document}

\title{Searching for Earth/Solar Axion Halos}

\author{
Abhishek Banerjee\affmark[$\oplus$], 
Dmitry Budker\affmark[$\odot$]\affmark[$\!\!\rightmoon$], 
Joshua Eby\affmark[$\oplus$], 
Victor V. Flambaum\affmark[$\leftmoon$]\affmark[$\!\!\odot$],\\
Hyungjin Kim\affmark[$\oplus$], 
Oleksii Matsedonskyi\affmark[$\oplus$],
and Gilad Perez\affmark[$\oplus$]
\vspace*{.2cm}\\
{\it\affaddr{\affmark[$\oplus$]\Weizmann}}\\
{\it\affaddr{\affmark[$\odot$]\Mainz}}\\
{\it\affaddr{\affmark[$\leftmoon$]}\sw}\\
{\it\affaddr{\affmark[$\rightmoon$]\Berkeley}}\\
}

\date{}
\maketitle

\begin{abstract}
We discuss the sensitivity of the present and near-future axion dark matter experiments to a halo of axions or axion-like particles gravitationally bound to the Earth or the Sun. Such halos, assuming they are formed, can be searched for in a wide variety of experiments even when the axion couplings to matter are small, while satisfying all the present experimental bounds on the local properties of dark matter. The structure and coherence properties of these halos also imply novel signals, which can depend on the latitude or orientation of the detector. We demonstrate this by analyzing the sensitivity of several distinct types of axion dark matter experiments.   
\end{abstract}
\newpage

\section{Introduction}

Axion-like particles (ALPs)\footnote{We will use the terms axion and ALP interchangeably in this work.} can form coherently oscillating classical field configurations.
Such an oscillating field behaves as non-relativistic matter from the early universe, and can account for dark matter (DM) in the present universe~\cite{Dine:1982ah,Abbott:1982af,Preskill:1982cy}. 

A variety of experiments have been proposed to search for such DM candidates.
Those experiments include axion haloscopes using microwave cavities~\cite{Du:2018uak, Asztalos:2003px, Asztalos:2009yp}, experiments exploiting magnetic resonance~\cite{Budker:2013hfa, Graham:2013gfa, Ruoso:2015ytk}, interferometry~\cite{Liu:2018icu, Nagano:2019rbw, Obata:2018vvr, DeRocco:2018jwe}, precision magnetometers and LC circuits~\cite{Sikivie:2013laa, Chaudhuri:2014dla, Kahn:2016aff, Silva-Feaver:2016qhh, Gramolin:2020ict}, atomic transitions~\cite{Sikivie:2014lha}, searches for an oscillating neutron electric dipole moment~\cite{Abel:2017rtm}, and many others (for a more complete list of axion dark matter detection experiments, see Refs.~\cite{Kim:2008hd, Graham:2015ouw} and references therein).
These experiments cover a wide range of ALP DM particle mass, from $10^{-22}\eV$ to $10^{-2}\eV$. 
Since all of them are based on the assumption that the dark matter is a coherently oscillating ALP field, the experimental reach depends on the local DM density. 

The DM density near the Earth cannot be measured directly through gravitational effects and is therefore subject to both observational and theoretical uncertainties. It is known that throughout the cosmological history, axions may form gravitationally-bound objects, whose density can be orders of magnitude larger than the local dark matter density. 
Typical examples include axion miniclusters~\cite{Hogan:1988mp,Kolb:1993zz} and boson stars~\cite{Kaup:1968zz,Ruffini:1969qy,Colpi:1986ye}  (see also~\cite{Arvanitaki:2019rax,Fonseca:2019ypl,Co:2019jts,Braaten:2019knj} for recent discussions).
Being much denser than the average galactic DM density, these small-scale objects could boost the discovery potential of aforementioned axion dark matter experiments, should they exist. 
To exploit a density enhancement, the encounter rate between Earth and these small-scale objects should be large enough such that one may expect several transient signals within an experimental timescale. 
However, the encounter rate is inversely proportional to the density of such objects, so even if bound objects make up the entirety of the DM in the galaxy,
one can only expect a mild density enhancement within a reasonable experimental timescale (see {\it e.g.}~\cite{Banerjee:2019epw} for a detailed discussion). It was recently pointed out that microwave atomic clocks, being sensitive to the mass range around $10^{-6}$~eV, may benefit from the density enhancement of transient boson stars \cite{Kouvaris:2019nzd}.
Boson stars with an axion-like potential, in the vacuum, can be generically divided into two distinctive phases, denoted as dilute and dense stars~(see for instance~\cite{Braaten:2019knj}).
In the dilute case, the field value is small, self interactions are negligible, and the axion self-gravity is balanced by the kinetic pressure, while the self interactions can be treated perturbatively~\cite{PhysRevD.84.043531, Eby:2018dat, Eby:2020eas, Schiappacasse_2018, PhysRevD.97.063012}.
One can also seek denser solutions that are subject to strong non-linear and relativistic effects~\cite{Levkov:2016rkk, Braaten:2016dlp, Panotopoulos:2019mis}, which are however unstable and might be short lived~\cite{Visinelli:2017ooc, Chavanis:2020hux, Muia:2019coe, Eby:2016cnq, 1806948}.
In this work we only consider the former possibility of dilute boson stars.

In~\cite{Banerjee:2019epw} the possibility of small-scale ALP objects (smaller than the size of solar system) becoming gravitationally bound to other astrophysical objects, such as the Earth or the Sun, was discussed from a phenomenological point of view. 
In the case of an Earth-bound halo, the strongest constraint is derived from lunar laser ranging~\cite{Adler:2008rq}, and for a Sun-bound halo, from planetary ephemerides~\cite{Pitjev:2013sfa}. The halos allow for a significant density enhancement, in principle as big as  twenty  and four orders of magnitude for the Earth and solar halo cases, respectively.  Despite these large enhancement factors, the densities 
of the terrestrial and solar halos are relatively small compared to those obtained in the dense phase of the boson stars, namely, they were found to be described effectively as dilute stars in the presence of a background potential~\cite{Banerjee:2019epw}. Furthermore, it is straightforward to verify that the perturbation from astrophysical objects, such as the moon on the surface of the Earth (in the case of the Earth halo), or the planetary motions (in the case of the Solar halo) are negligible and would not distort the halo configuration.   

In this work, we do not discuss the question of formation of these objects or  assume a specific value for the DM-halo density. 
What is the dark matter density within our solar system is a complicated astrophysical question and a subject of active research. Instead, we describe the state of the art regarding this issue.
There has been debate in the literature about whether DM can be efficiently captured by the Sun, leading to a large overdensity in the solar system. 
One can show that elastic two-body scattering of light bosons against the Sun is unlikely to give rise to a large overdensity, as the scattering rate is low and typical energy transfer of the DM is small compared to its initial energy.
Reference \cite{Xu:2008ep} considered three-body interactions involving the Sun, planets, and DM particles, concluding that this process can be efficient enough to give rise to large overdensities. By the authors' estimation, the DM density at the position of the Earth would be close to $10^{-20}$ g/cm$^3$, about four orders of magnitude higher than the standard estimate of $\rho_{\rm local} = 0.4$ GeV/cm$^3 = 7\times10^{-25}$ g/cm$^3$. Interestingly, the predicted density does not depend on the DM particle mass $m_\phi$ and  is of the same order as the gravitational constraint from solar system ephemerides \cite{Pitjev:2013sfa}.
In \cite{Khriplovich:2009jz, Khriplovich:2010hn}, the authors consider the same processes but point out large uncertainties in the estimation; they conclude that in the most optimistic scenario an overdensity nearly as large as that of \cite{Xu:2008ep} is possible, though the authors caution that this might be an overestimation of the actual effect. 
In a subsequent comment~\cite{Edsjo:2010bm}, it was claimed that the above work did not take into account the inverse process of three-body ejection of DM particles, which would counterbalance the capture rate found previously.
However, it was discussed in another comment~\cite{Khriplovich:2010ng} that the time scale for the inverse process is subject to a large uncertainty. 

It is relevant to mention yet earlier work, in the context of electroweak mass DM (i.e., weakly interacting massive particles, WIMPs), where simulations of capture in the solar system have been performed (including both capture and ejection processes in two-body and three-body interactions)~\cite{Lundberg:2004dn, Peter:2009mi, Peter:2009mm}, which resulted in much smaller densities.  
In the context of light scalar DM, this discussion appears inconclusive, as the detailed simulations assume  WIMP-nucleon cross sections, and there remains a possibility that there is some light scalar-specific dynamics that distinguishes it from the WIMP case. For example, galaxy-scale simulations of formation of light-scalar DM halos in the presence of background gravitational potentials (modelling the baryonic components of galaxies) are recently being performed~\cite{Veltmaat:2019hou}. 
There remain significant limitations in these simulations (e.g. they assume a fixed baryonic potential, ignoring backreaction from the DM); however, a preliminary conclusion of their study is that DM can become captured and bound to the central potential, resulting in a central DM structure whose profile closely resembles a `gravitational atom’.
This is interesting for our current work because, though the simulation of~\cite{Veltmaat:2019hou} are conducted in the regime of ultralight DM (where $m_\phi\sim 10^{-22}$ eV), the structure of the bound DM matches what we predict for Earth and Solar axion halos. It is thus possible that the same scalar field relaxation processes observed in~\cite{Veltmaat:2019hou}, which consistently give rise to solitons in DM-only simulations~\cite{Schive:2014dra, Schive:2014hza, Mocz:2017wlg, Eggemeier:2019jsu}, might give rise to bound halo-like structures at solar system scales when external gravitational influences are present (see also~\cite{anderson2020direct, Vaquero:2018tib, Fairbairn:2017sil} for additional recent studies).

In summary, the ALP DM density at the surface of the Earth may be significantly enhanced, compared to what is commonly assumed; however, the precise value is model dependent and is subject to large uncertainties.
We therefore believe that this possibility merits a study, which is the focus of this work. We point out that the experimental signatures of a bound axion halo do not just depend on the density enhancement, but also on an increased coherence time and  a modification of spatial gradients in the DM field, as compared to the virialized DM case.

The paper is organized as following.
In Sec.~\ref{sec:halo}\,, we summarize the main properties of the axion halos, based on the results previously derived for a dilaton-like (scalar) field halo~\cite{Banerjee:2019epw}.
We also highlight novel features and unique signatures of this scenario, by comparing to the standard local-DM scenario.
In Sec. \ref{sec:Experiments}\,, we analyse the sensitivity of the experiments testing the couplings of the axion field to the Standard Model. In Sec. \ref{sec:MaxMass}\,, we reinterpret sensitivity to QCD axion parameters as a probe of the mass of the axion halos in the solar system, bound by the Earth or the Sun. We conclude in Sec. \ref{sec:conclusion}\,.
We work in natural units, where $\hbar = c = 1$, which in this case can be done without loss of generality~\cite{Antypas:2019yvv}.

\section{Axion Halo Properties} \label{sec:halo}

We consider compact astrophysical objects composed of ALPs. 
In particular, we focus on scenarios where such objects are gravitationally bounded to the Earth or the Sun; we call such an object a {\it Solar or Earth axion halo}.
The axion halo we are considering here shares similar properties with usual boson stars (see {\it e.g.}~\cite{Kaup:1968zz,Ruffini:1969qy,Colpi:1986ye,Chavanis:2011zi,Eby:2018dat}), except that it is bounded by a gravitational potential of an external body rather than its own self-gravity. A consistent mechanism for relaxation of scalar field density onto external bodies remains a topic for a future study.
Throughout this work, we will assume that such halo formation is possible, and in lieu of a concrete prediction for the density of the halo, we will bracket the allowed range for this parameter; we also assume for simplicity that the center of the axion halo coincides with the center of the external host body. 
In this section, we review some properties of axion halo, following and expanding on the previous work on scalar field halos~\cite{Banerjee:2019epw}. 

We consider either the Earth or the Sun as a host of an axion halo.
As the axion halo is maintained by the gravitational potential of the host, its mass $M_\star$ is assumed to be smaller than $M_{\rm ext}(R_\star)$, defined as the host mass that is enclosed within the radius $R_\star$ of the axion halo.
For clarity we 
identify the maximal axion halo mass
\bea \label{mstar} 
M_{\star}^{\rm max} \equiv {\rm min}\left[M_{\rm max}, \frac{M_{\rm ext}(\Rs)}{2}\right],
\eea
where $M_{\rm max}$ is the maximally allowed mass for axion halo that is consistent with local measurements of gravity and solar system ephemerides~\cite{Adler:2008rq,Pitjev:2013sfa,Banerjee:2019epw}.
A factor of two is introduced to ensure that the object is maintained due to the gravitational potential of the external body rather than self-gravity.

To approximate the shape of the density profile, we used the procedure outlined in \cite{Banerjee:2019epw}: the axion halo profile is approximately exponential at $\Rs \gg R_{\rm ext}$; it is Gaussian at $\Rs \ll R_{\rm ext}$; and in the intermediate region we interpolate between the two.
For example, when $R_\star \gg R_{\rm ext}$, the axion field has an exponentially decaying profile with the distance from the center of the host body $r$, and oscillating in time with a frequency approximately equal to the axion mass $m_\phi$ in the halo rest frame:
\bea\label{eq:phiprofile}
\phi(r,t) \simeq \cos [m_\phi t + \theta(r,t) ] \exp \left(- {r}/{\Rs}\right) & \,\, \text{for}\,\, \Rs \gg R_{\rm ext},
\eea
where $\theta(r,t)$ is a phase factor slowly varying with respect to the position and the time.
The time scale over which the phase factor changes by an order one value is the coherence time scale that we discuss below. 
The corresponding energy density is given by $\rho_\star \simeq m_\phi^2 \phi^2/2$.  
Since the time dependence is given by the same factor $\cos\left(m_\phi t\right)$ throughout, for the remainder of this work we will no longer write it explicitly and refer to $\phi$ as the space-dependent part only.

The size of the halo can be determined as a balance between the repulsive gradient energy, {$U_{\rm grad} = \int d^3x (\nabla \phi)^2 \sim M_\star / (m_\phi^2 R_\star^2)$}, and the attractive gravitational potential energy, ${U_{\rm grav} \sim GM_{\rm ext} M_\star/R_\star}$. The resulting radius is
\bea \label{eq:Rstar}
\Rs \simeq \frac{1}{G  M_{\rm ext}(\Rs) m_\phi^2 },
\eea
which is independent of $M_\star$.
That is, the axion halo mass is a free parameter; later, in the estimations of experimental sensitivities, we will set its maximum value by Eq.~\eqref{mstar} and quantify its effect for values smaller than $M_\star^{\rm max}$.  
For $\Rs > R_{\rm ext}$, the radius scales as $\Rs \propto m_\phi^{-2}$, while for $\Rs< R_{\rm ext}$, it scales as $\Rs \propto m_\phi^{-1/2}$ since the enclosed external mass scales as $M_{\rm ext}(R_\star) \propto R_\star^3$.
If the axion halo profile extends to sufficiently large radii, experiments on Earth's surface will benefit from a large axion halo density. 
For an Earth-based halo, the relevant requirement is $\Rs \gtrsim R_\oplus$, implying $m_\phi \lesssim 10^{-9}$ eV, while, for a Sun-based halo, we require $\Rs \gtrsim 1$ AU which implies $m_\phi \lesssim 10^{-14}$ eV.
Note that the radius coincides with the de-Broglie wavelength $\lambda_{\rm dB}  = (m_\phi v_\star)^{-1}$, where $v_\star \simeq \sqrt{G M_{\rm ext}/\Rs}$.

\begin{figure}
\centering
\includegraphics[width=0.48\textwidth]{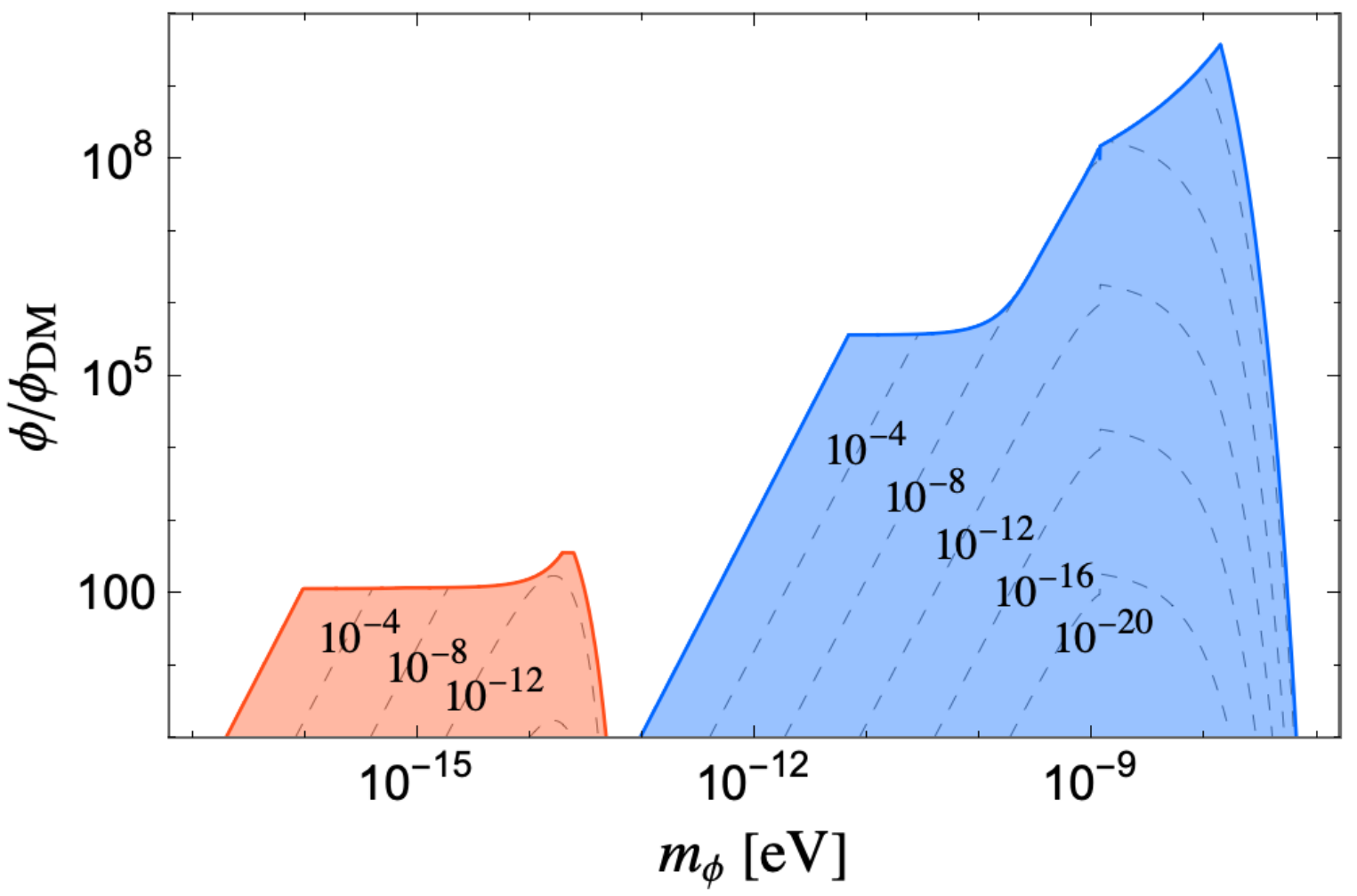}
\hspace{.2cm}
\includegraphics[width=0.48\textwidth]{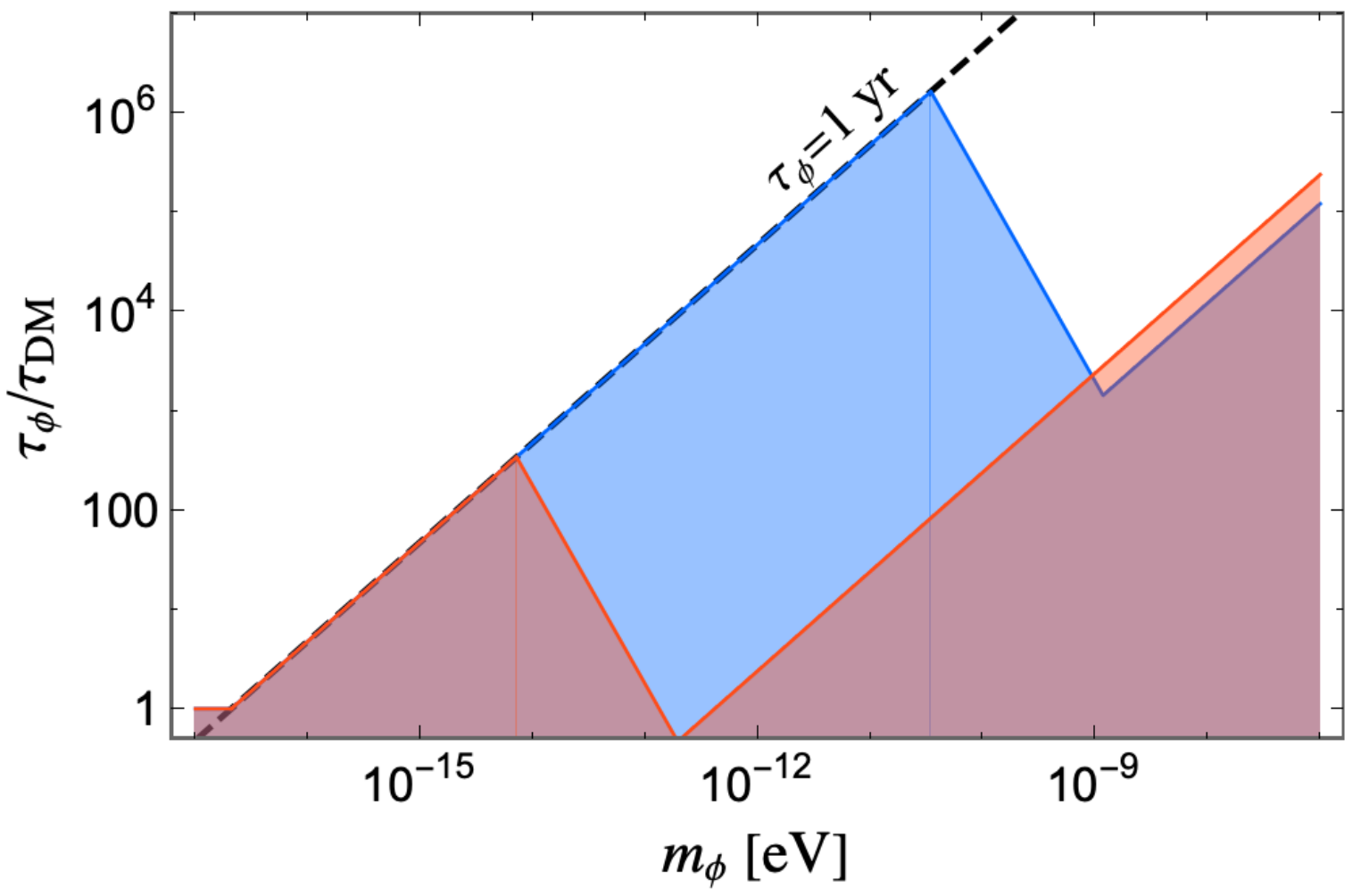}
\caption{ 
Modifications in the axion halo scenario compared to the background DM case.
 Left:  
 Modification to the field value for the Earth halo (blue) and solar halo (red) compared to the usual ALP DM case. Solid lines correspond to maximal halo mass $M_\star=M_\star^{\rm max}$, given by Eq.~(\ref{mstar}), which is currently allowed by gravitational constraints \cite{Adler:2008rq,Pitjev:2013sfa} (see \cite{Banerjee:2019epw} for further details). Dashed lines correspond to the halos with a mass smaller than the maximal, with contours indicating the halo mass as a fraction of the Earth or the Sun mass. 
 Right:  
 Modification to the coherence time for the Earth halo (blue) and solar halo (red). For estimations of experimental sensitivity, we limit the maximal data acquisition time to one year.}
\label{fig:enhancement}
\end{figure}

Having determined the radius, the density of axion halo is given as $\rho_\star = 3 M_\star / 4\pi R_\star^3 \simeq m_\phi^2 \phi^2 /2 $, and thus, the field amplitude inside the axion halo is $\phi \simeq \sqrt{2\,\rho_\star}/m_\phi$. In the left panel of Figure~\ref{fig:enhancement}, we consider axion halo either hosted by Earth (Earth halo) or Sun (solar halo), and compare the field amplitude of these halos to that of the standard axion DM scenario.   
The outer envelopes of the red and blue regions represent the maximal axion halo mass allowed, given in Eq.~\eqref{mstar}; the contours represent smaller values of $M_\star/M_{\rm ext}$.
Note that the standard axion DM scenario is characterised by a local field value $\phi_{\rm DM}$, coherence time $\tau_{\rm DM}$, and virial velocity $v_{\rm vir}$ given by
\bea
\phi_{\rm DM} \simeq \sqrt{2 \rho_{\rm local}}/m_\phi\,, \quad 
		\tau_{\text{DM}} \simeq 1/m_\phi v_{\rm vir}^2\,, \quad 
		v_{\rm vir} \simeq 10^{-3},
\eea 
where $\rho_{\rm local}=0.4\GeV/\rm{cm}^3$ is the local density of the background axion dark matter (in our numerical calculations we took $v_{\rm vir} = 230$ km/sec $=7.7\times10^{-4}$).

The coherence time of the axion halo requires a more careful discussion.  
If the axion halo is an exact ground-state solution of the equation of motion (a condensate), as is often assumed about self-gravitating axion stars \cite{Ruffini:1969qy}, its oscillations remain coherent on infinite timescales.  
If, on the other hand, the axion halo is not the true ground state solution, 
but is a collection of axions with some distribution of velocities centered around $v_\star$, then we can describe the axion halo as a sum of incoherent subcomponents with slightly shifted oscillation frequencies and phases.
This picture is potentially motivated by simulations of axion star formation which observe lingering fluctuations on scales larger than $\l_{\rm dB}$ \cite{Eggemeier:2019jsu} (see also \cite{Centers:2019dyn} for the phenomenological implication of the picture for the large-scale bosonic DM halo). 
In this case, the coherence timescale of the axion halo can naively be estimated as $\tau_{\phi} = (m_\phi\,v_\star^2)^{-1} = m_{\phi}\Rs^2$, which is,
\bea
 \tau_{\phi} \simeq 10^3 \,{\rm sec} \times \max\left[1, \Big(\frac{10^{-9} \eV}{m_\phi} \Big)^3 \right]
\eea 
for the Earth-based axion halo, whereas for Sun-based axion halo we find
\bea
\tau_{\phi} \simeq 10^{10}\, {\rm sec} \,
	(10^{-15}\eV/m_{\phi})^3\, 
\eea
as long as $m_{\phi} < 10^{-13}\eV$ ({\it i.e. }the halo radius is greater than the radius of the Sun). We will use this estimation for the coherence time of the axion halo, as we expect it represents a conservative lower bound on the true coherence time, though a detailed discussion of the condensate picture is deserving of more careful study in the future.
In the right panel of Figure~\ref{fig:enhancement}, we show the 
modification to the coherence time in the axion halo $\t_{\phi}$ relative to the standard halo model result $\t_{\rm DM}$. 

We choose to work in the basis where axion couples to Standard Model only derivatively except for anomalous couplings to gauge bosons.
For the experiments relying on derivative couplings, the sensitivities depend on the spatial gradient of the axion field. 
This gradient generically has two distinct components $\nabla\phi = \nabla_r\phi + \nabla_\perp\phi$, which we refer to as the radial and the tangential components respectively, labelled by their orientation with respect to coordinates centered at the Earth or the Sun.

The radial component comes from the variation of the radial field profile, Eq.~\eqref{eq:phiprofile}, which is given by
\bea
\frac{|\nabla_{r} \phi|}{\phi} = \frac{1}{\Rs}.
\eea
This is a static gradient, oriented along the radial direction; for convenience it can be written in terms of $v_\star = (m_\phi\,\Rs)^{-1}$, though this should be interpreted as a velocity dispersion rather than a net flow of axions (see relevant discussions in~\cite{Stadnik:2013raa,Tan:2018jar}). This makes sense at the level of the classical equation of motion, where in the rest frame of the halo the kinetic energy in the $\phi$ field is $|\nabla_r\phi|^2/(2\,m_\phi\,\phi^2) = 1/(2\,m_\phi\,\Rs^2) = m_\phi\,v_\star^2/2$.

\begin{figure}
\centering
\includegraphics[width=0.48\textwidth]{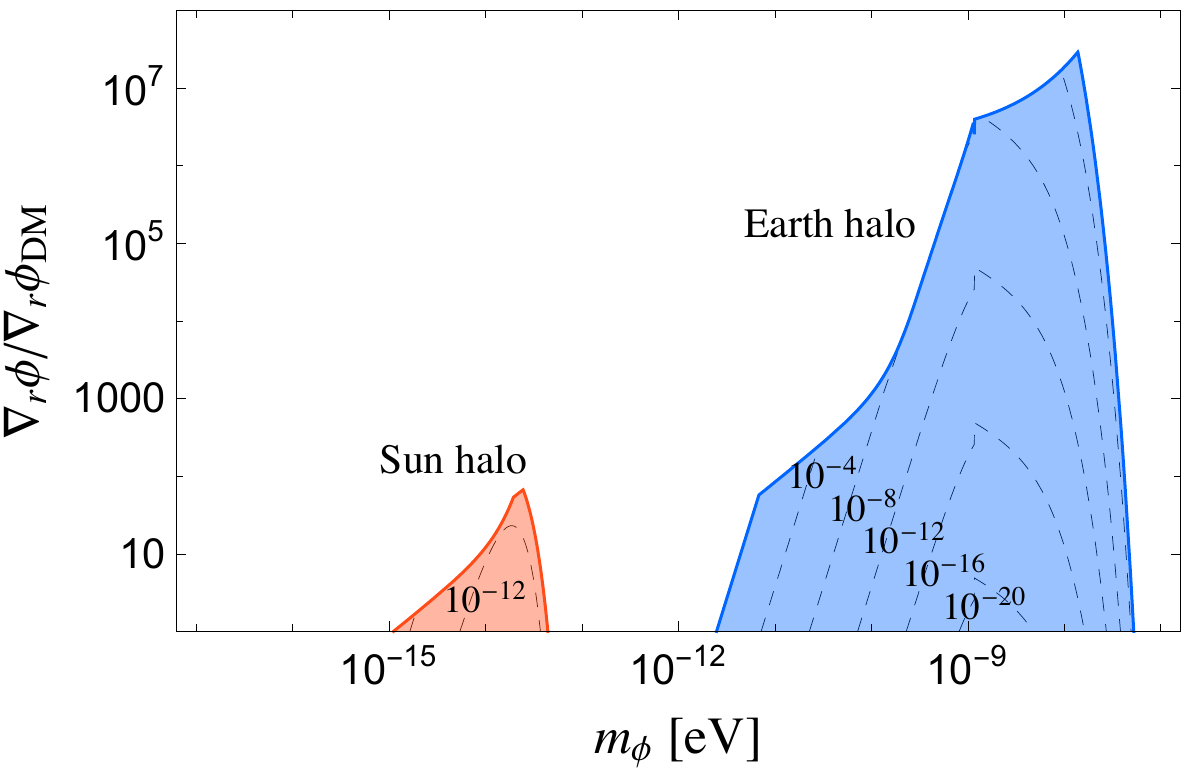}
\hspace{0.2cm}
\includegraphics[width=0.48\textwidth]{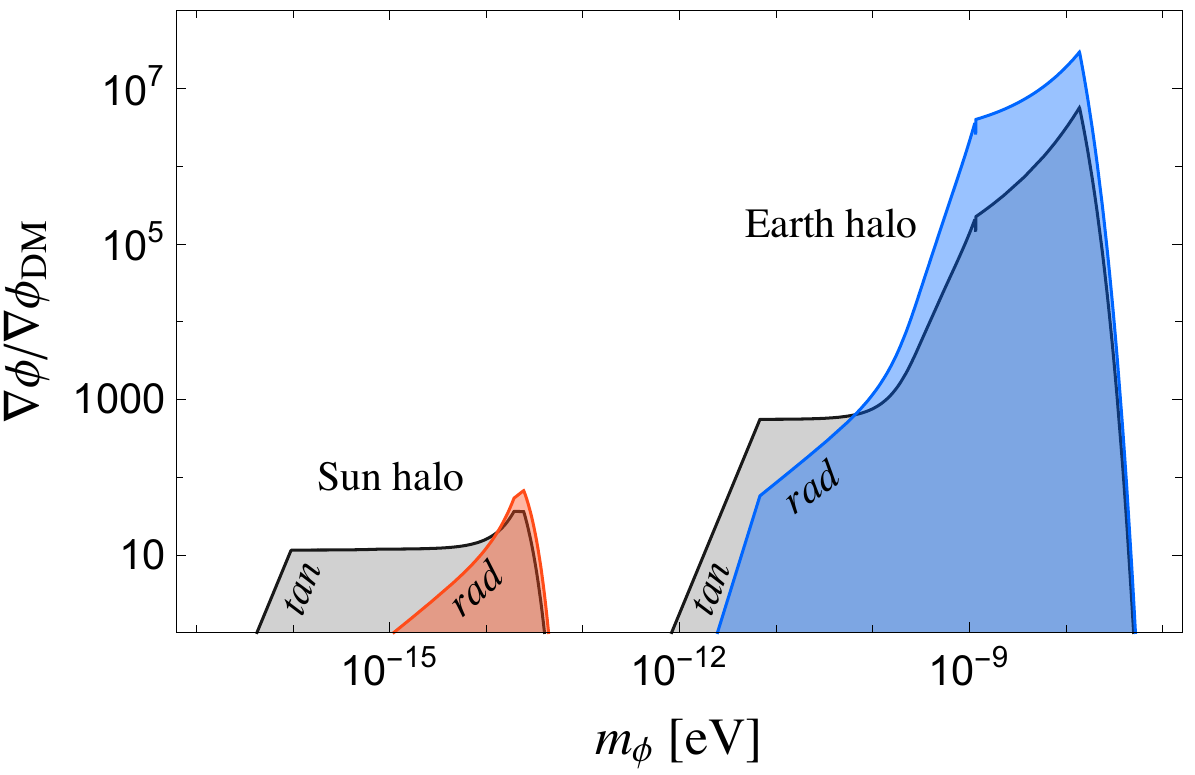}
\caption{Left:  
Modification to the radial field gradient for the Earth halo (blue) and Sun halo (red) with respect to the ALP DM case. Solid lines correspond to the maximal halo mass allowed by the current constraints, which is given by Eq.~(\ref{mstar}). Dashed lines correspond to halos with a mass smaller than maximal, with contours indicating the halo mass as a fraction of the host mass (Earth or Sun). 
Right: Comparison between the radial (red and blue) and the tangential (grey) gradient modifications.}
\label{fig:gradenhancement}
\end{figure}

The tangential component of the gradient appears as a consequence of the relative motion between the halo and an experimental device. The corresponding field gradient is 
\bea
\frac{|\nabla_{\perp} \phi|}{\phi} =m_\phi v_{\text{rel}}  \,,
\eea
where $v_{\text{rel}}$ is approximately the speed of a device in the rest frame of halo. Its value depends on the halo angular momentum, which is a free parameter at the level of this analysis. 
If the halo is non-rotating,  
we can take $v_{\text{rel}}$ equal to the Earth orbital speed around the Sun $v_{\rm rel}^\odot \sim 10^{-4}$ (for the solar halo) or to the Earth surface speed, $v_{\rm rel}^\oplus \sim 10^{-6}$ at the equator (for the Earth halo).  
For simplicity we assume zero angular momentum for the axion halo, so that the maximum tangential gradient for an experimental apparatus is of order the optimistic estimate given here.  

In Figure~\ref{fig:gradenhancement}, we present the 
modification to the field gradient as a function of ALP mass, relative to the standard background DM result.
In the left panel, we illustrate the radial component for different choices of $M_\star$ (the contours correspond to $M_\star/M_{\rm ext}$), while, in the right panel, we compare the radial (red and blue) and tangential (grey) gradients. The tangential gradient becomes most important at small values of $m_\phi$, where the radial gradient $1/\Rs$ is suppressed. This implies that the position and orientation of the device can be important to determine the signal strength.

\subsection{Detector Orientation, Latitude, and Modulation of Signals} \label{app:modulation}

Depending on the spatial orientation of an experimental device, it may be sensitive to either radial or tangential (or both) components of the gradient, and the amplitude of the signal may modulate over the course of a day or year. For the Earth halo, both gradient components are constant in time; the radial gradient can be maximized by choice of orientation of the experimental apparatus with respect to the surface of the Earth, whereas the tangential gradient depends in addition on the latitude of the detector, as the relative velocity $v_{\rm rel}^\oplus$ is faster at the equator and goes to zero at Earth's poles. 
For the solar halo, both gradient components modulate strongly with a period of one solar day, with a weaker additional annual modulation, because as the Earth rotates and revolves around the sun, the orientation of the detector rotates as well. 

Suppose for concreteness that an experiment is searching for a signal proportional to $\vec{r}_{\rm det}\cdot\nabla\phi$, where $\vec{r}_{\rm det}$ (denoted as the ``detector orientation" in what follows) is some vector quantity in the experiment ({\it e.g.} it is related to the orientation of nuclear spin in a nuclear magnetic resonance experiment).  
Though we focus on the example of a dot product, it is simple to extrapolate to the case where the signal is given by the cross product $\vec{r}_{\rm det}\times\nabla\phi$ instead.

In the case of an Earth-based halo, both components of the gradient are constant with time. The radial gradient will be maximized when $\vec{r}_{\rm det}$ is oriented perpendicular to the surface of the Earth, whereas the tangential gradient is maximized when $\vec{r}_{\rm det}$ is instead parallel to the surface of Earth, oriented along the direction of Earth's rotation about its axis\footnote{Note that if the signal is proportional to a cross product $\vec{r}_{\rm det}\times\nabla\phi$, then it is possible to maximize both components of the gradient, by orienting $\vec{r}_{\rm det}$ both parallel to the surface of the Earth and perpendicular to direction of Earth's rotation.}. Further, the latter depends on the latitude of the detector, because the linear speed of Earth's rotation is maximized at the Equator and minimized at the poles. If $\vec{r}_{\rm det}$ is pointed along the lines of latitude, 
then we can parametrize the dependence on the latitude $\ell$ of the detector simply as
\begin{equation}
 \hat{r}_{\rm det} \cdot \left(\frac{\nabla_\perp\phi}{(m_\phi\,v_{\rm rel})\,\phi}\right)  = \cos\ell
 		\qquad {\rm (Earth\,halo)},
\end{equation}
where the hat denotes a unit vector. Note that in our notation, a latitude of $x^\circ$N is denoted by $\ell=+x$ and $y^\circ$S is denoted by $\ell=-y$.

\begin{figure}
\centering
\includegraphics[width=0.7\textwidth]{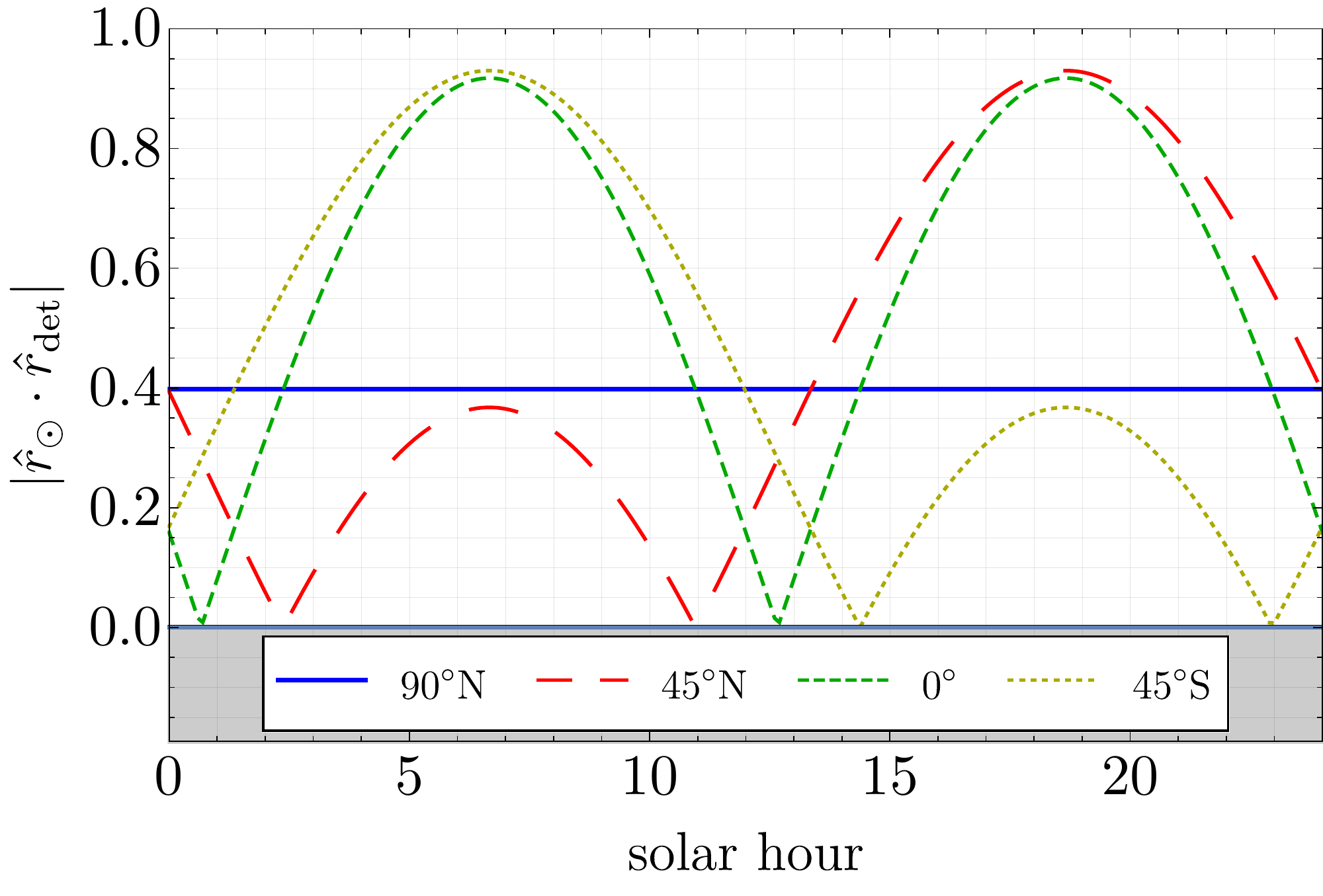}
\caption{The magnitude of radial gradients in solar halo scenario, normalized so that the maximum possible signal corresponds to unity. The blue thick, red long-dashed, green short-dashed, and yellow dotted lines correspond to detector latitude of $\ell = 90^\circ$N, $45^\circ$N, $0^\circ$, and $45^\circ$S (respectively).}
\label{fig:radgradient}
\end{figure}

The case of the solar halo is more complex. As before, the effect depends on the latitude of the detector, but as the Earth rotates and revolves around the sun, a detector oriented along $\hat{r}_{\rm det}$ will also see modulation of the signal along both the radial and tangential directions. To analyse this, we use the results of \cite{McCabe:2013kea} which determine the position and orientation of the Earth and Sun as functions of time.

The radial gradient varies with the product $\hat{r}_{\rm det}\cdot \hat{r}_\odot$, where $\hat{r}_\odot$ is a unit vector pointed from the Sun to the Earth. In Figure \ref{fig:radgradient}, we illustrate the solar-daily modulation of the signal at $\ell = 90^\circ$, $45^\circ$, $0^\circ$, and $-45^\circ$ (blue thick, red dashed, green long-dashed, and yellow dotted, respectively). The signal from radial gradient does not change on different days of the year, as the axion halo is spherically symmetric and Earth's orbit is approximately circular. The curves are normalized such that the quantity on the vertical axis approaches unity when the two vectors are perfectly aligned, though we do not achieve such at the latitudes depicted; the maximum is reached at some point in the day if the detector is located at a latitude of roughly $23^\circ$N or $23^\circ$S, corresponding to the obliquity of the ecliptic (the tilt of the Earth's rotation axis).

\begin{figure}
\centering
\includegraphics[width=0.48\textwidth]{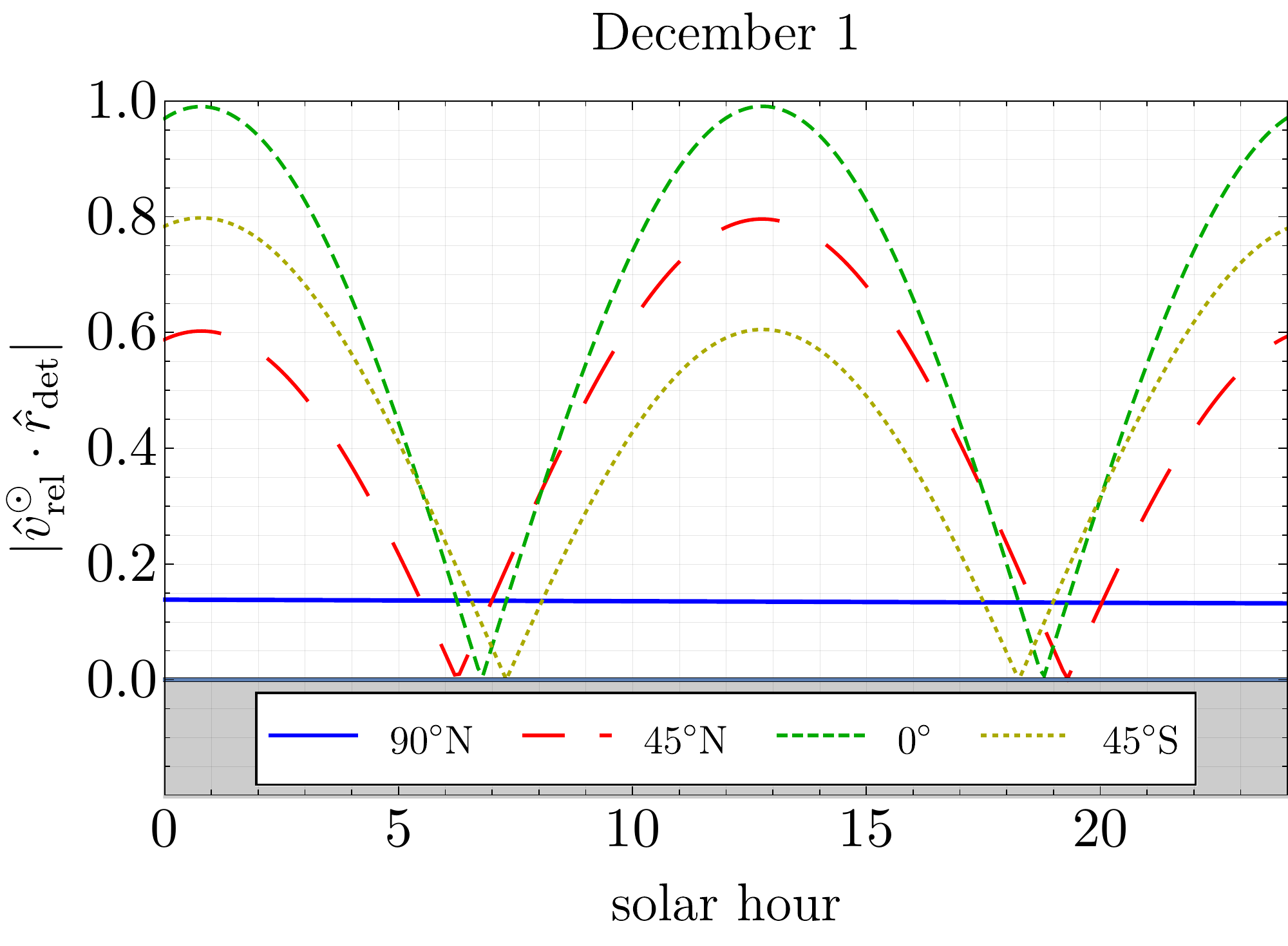}
\hspace{0.2cm}
\includegraphics[width=0.48\textwidth]{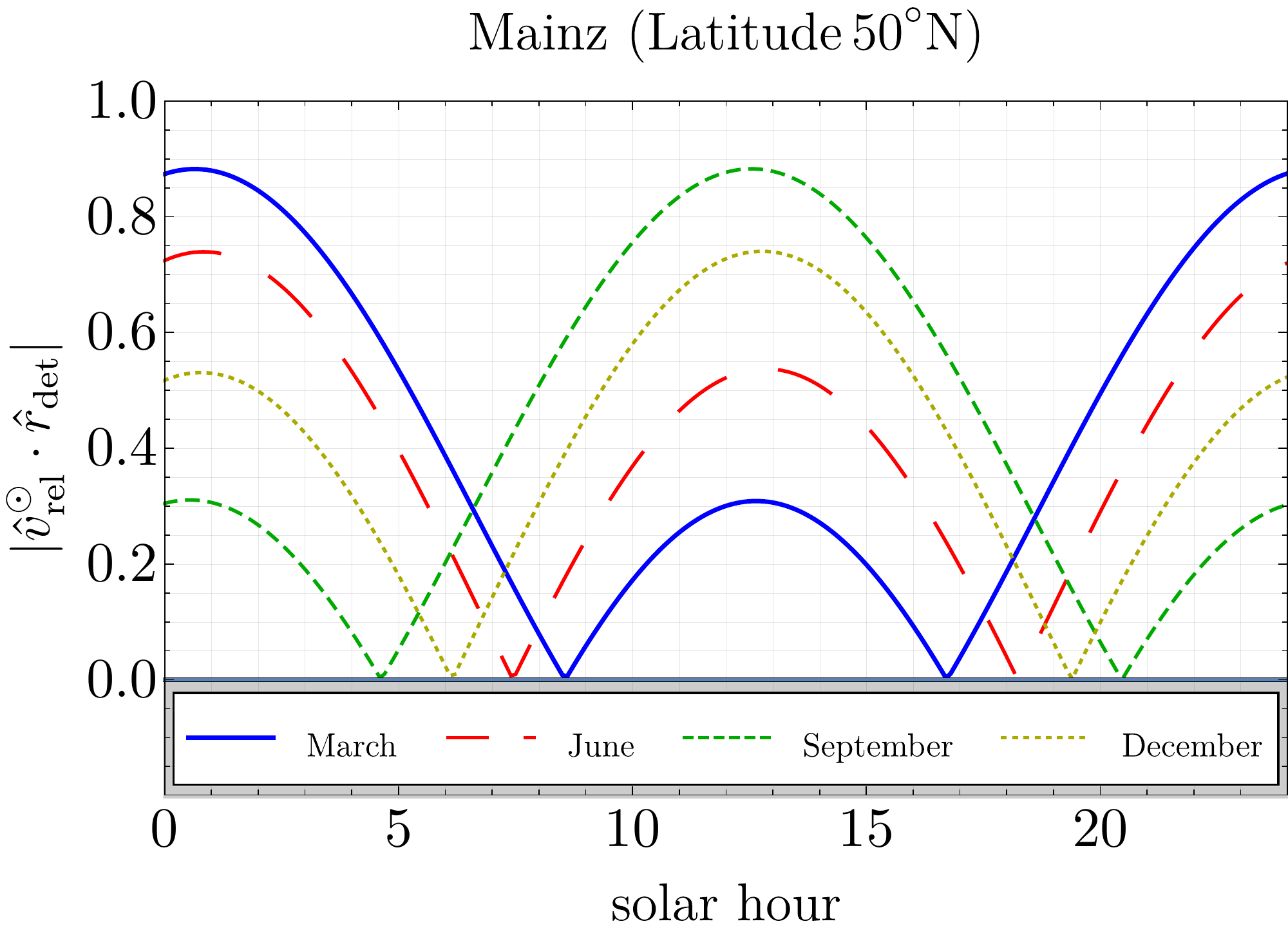}
\caption{The magnitude of tangential gradients in the solar halo scenario, normalized so that the maximum possible signal corresponds to unity. In the left panel, the time is fixed to December 1 and the blue thick, red long-dashed, green short-dashed, and yellow dotted lines correspond to detector latitude of $\ell = 90^\circ$N, $45^\circ$N, $0^\circ$, and $45^\circ$S (respectively); in the right panel, the latitude is fixed to $50^\circ$N (Mainz, Germany) and the same set of colored lines correspond to the signal in March, June, September, and December (respectively).}
\label{fig:tanggradient}
\end{figure}

The tangential gradient for the solar halo varies similarly with latitude $\ell$, but also with time of year. In Figure \ref{fig:tanggradient}, we illustrate this by displaying the signal strength, proportional to $\hat{v}_{\rm rel}\cdot \hat{r}_{\rm det}$, in two cases: in the left panel, at a fixed time of year (December 1) at latitudes $\ell = 90^\circ$, $45^\circ$, $0^\circ$, and $-45^\circ$; and in the right panel, at fixed latitude ($50^\circ$N, the location of Mainz, Germany) on the first day of March, June, September, and December. The lines in both panels are blue thick, red long-dashed, green short-dashed, and yellow dotted, respectively.

The existence of separate radial and tangential gradients is a unique prediction of the axion halo model. This implies a method to distinguish a bound axion halo from background DM by using directional information, as the detector orientation and its location can modify the signal; this is possible even if the axion halo has the same mass density as the background DM. We see further that in the solar halo case, there are additional oscillations of the signal with well-defined period of one solar day (note that (1 day)$^{-1} \simeq 10^{-5}$ Hz). Therefore in a Fourier analysis of the oscillating signal, the halo scenario predicts a sideband corresponding to an axion mass of roughly $10^{-19}-10^{-20}$ eV. The additional annual modulation in the tangential gradient corresponds to a sideband at $10^{-7}$ Hz, or $10^{-22}$ eV.

\section{Experimental Probes} \label{sec:Experiments}
We demonstrate 
the experimental sensitivity to an axion halo by investigating a set of ongoing or proposed axion DM experiments. 
Among many other axion DM searches, we choose  experiments sensitive to the axion mass range $10^{-16}\eV < m_\phi < 10^{-6}\eV$,
a range in which the axion halo could be sufficiently large such that these searches could exploit a large density enhancement relative to the case of background DM. 
We evaluate the reach on coupling constants for these searches, and illustrate the main differences between the axion halo case and the axion DM scenario.

\subsection{ALP Search using nuclear magnetic resonance} 
We begin with ALP DM searches using nuclear magnetic resonance (NMR) techniques. 
From the UV perspective all the relevant interactions stem, to the leading order, from a single operator $(\phi/f) G\widetilde{G}$.  Since the zero mode of axion introduces CPV interactions in the chiral Lagrangian, it gives rise to nucleon electric dipole moment, for instance, through a pion loop diagram with an insertion of CPV nucleon-pion vertex. 
In addition, the CPV nucleon-pion vertex also induces CP-violating nucleon-nucleon interaction via pion exchange. The relevant interactions are
\bea \label{CASPEr}
{\cal L} \supset g_{\phi NN}\partial_\mu \phi \bar{N} \gamma^\mu \gamma_5 N  - \frac{i}{2} g_d\,\phi\, \bar{N} \sigma_{\mu\nu} \gamma_5 N F^{\mu\nu}+...,
\label{pheno}
\eea
where $N$ is the nucleon field, $F_{\mu\nu}$ is the electromagnetic field strength tensor, $g_{\phi NN}$ and $g_d$ are the coupling constants, and the dots correspond to other terms, for instance the nucleon-nucleon contact term. 

The first term in Eq.~\eqref{CASPEr} represents, for example, the coupling between axion and nuclear spin, while the second term is the axion-dependent 
nucleon electric dipole moment (EDM). 
These operators of the effective nuclear action produces CP-violating nuclear Schiff moment which actually induces atomic  EDM $d_a$ which is measured in experiments~\cite{Stadnik:2013raa}.
According to the Schiff theorem, nuclear EDM is screened by atomic electrons \cite{PhysRev.132.2194}. Atomic EDM is actually produced by a higher order electric multipole called the nuclear Schiff moment~\cite{Flambaum:1984fb}.
In atoms with zero electron angular momentum, the atomic EDM $\vec {d}_a$ and magnetic moment  $\vec {\mu}_a$ are directed along the nuclear spin $\vec{I}$. Therefore, the  effective interaction of the nuclear spin with external electric and magnetic fields may be presented in the following form:   
\bea
H \simeq - [(d_a/I)\vec{I} \cdot \vec{E} + (\mu_n/I)\vec{I} \cdot \vec{B}_\phi] \cos (m_\phi t)\,,
\label{eff_Ham}
\eea
where $\mu_n$ is the nuclear magnetic dipole moment. The axion-dependent atomic EDM and the effective magnetic field are (respectively) 
\bea
\vec{d}_a &=& g_{ad} \,\phi \,\vec{I}\,, \nonumber
\\
\vec{B}_\phi &=& g_{\phi NN} \gamma_n^{-1} \nabla \phi\,.
\eea
where $\gamma_n$ is the gyromagnetic ratio of a given nucleus.
We simply express $g_{ad} =\epsilon_{s}\, g_{d}$ with $\epsilon_s = 10^{-2}$, following~\cite{Budker:2013hfa}. Note that the gyromagnetic ratio cancels in Eq.~\eqref{eff_Ham} (since $\mu_n \propto \g_n$), so that the ALP-induced force to the nuclear spin is in fact independent of $\gamma_n$.

The Cosmic Axion Spin Precession Experiment (CASPEr) has been proposed to probe the above interactions using NMR techniques~\cite{Graham:2013gfa,Budker:2013hfa}; we briefly review the concept of this experiment here. 
We first discuss how CASPEr can probe the effective axion-induced atomic EDM (CASPEr-Electric).
A polarized material is placed in the external magnetic field background, $\vec{B}_{\rm ext}$, while its direction is aligned along the direction of the polarization. 
An external electric field $\vec{E}$ is also applied orthogonal to the direction of external magnetic field. 
Basically the atom has an axion-induced EDM, thus, the electric field exerts a torque to nuclear spins, which develop a non-zero angle with respect to the external magnetic field, and hence, a non-zero transverse magnetization is obtained.

If the spin precession frequency, {\it i.e.} the Larmor frequency $\Omega =2 \mu_n B_{\rm ext}$, matches to the frequency of the oscillating EDM $\omega \simeq m_\phi$,
the amplitude of the transverse magnetization is resonantly enhanced as 
\bea
M_\phi \simeq n \,p \,\mu_n d_a |(\vec{I}/I) \times \vec{E}| \, t,
\label{M_electric}
\eea
where $n$ is the nuclear spin density, $p$ is the material polarization. 
The halo profile $\phi(r)$ is to be evaluated at the position $r$ in the halo where the experiment takes place, {\it e.g.} $r=R_\oplus$ at the Earth's surface as measured from its center. The magnetic flux induced by this transverse magnetization is picked up by a pickup loop coupled to a superconducting quantum interference device (SQUID). 
The time $t$ over which a transverse magnetization can develop will be limited by one of a few timescales: the transverse spin relaxation time $T_2$, a property of the nuclear target; the coherence time scale of the axion field $\tau_{\phi}$; or the shot time $t_s$, defined as the time the experiment spends probing a given mass $m_{\phi}$ with a given strength of magnetic field.  We refer the reader to Appendix~\ref{app:CE} and to the original Refs.~\cite{JacksonKimball:2017elr,Garcon:2017ixh,Budker:2013hfa,Graham:2013gfa} for more details on the detection technique and bound extraction from sensitivity estimate. 

\begin{figure}[t]
	\centering
	\includegraphics[width=0.49\textwidth]{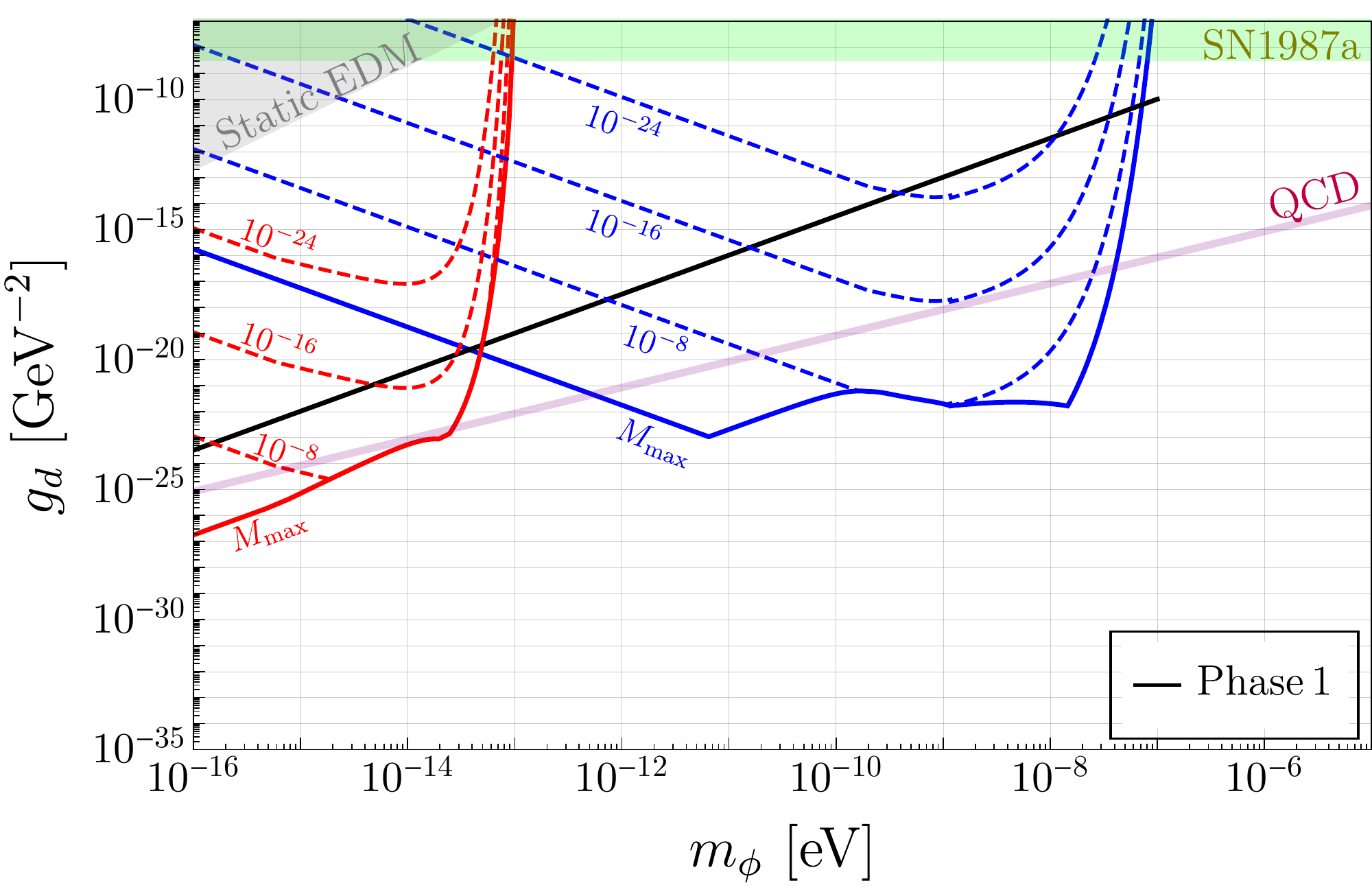} \,
	\includegraphics[width=0.49\textwidth]{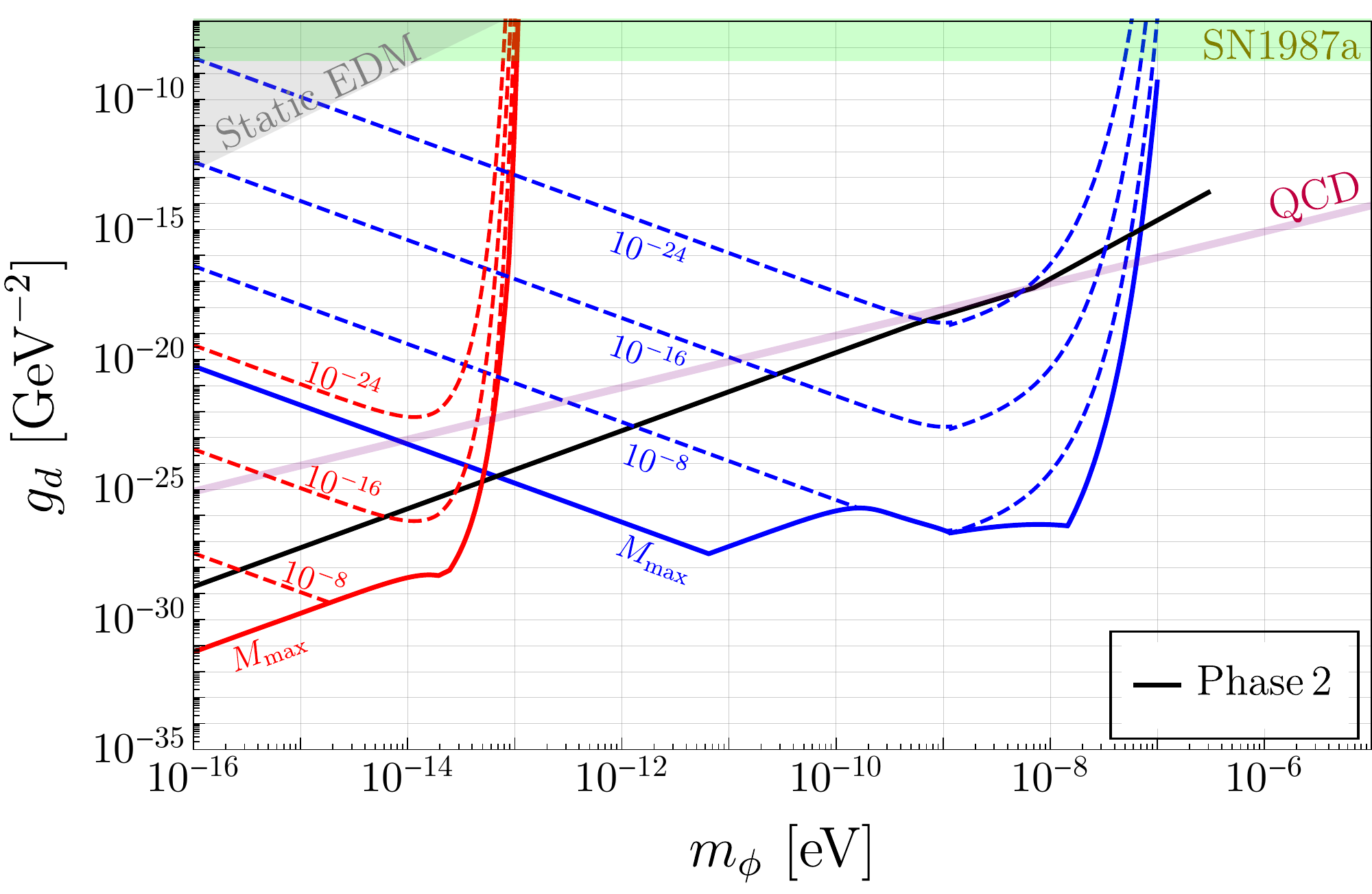}
	\caption{Sensitivity to $g_d$ in presence of an axion halo for CASPER-Electric; the thick blue (red) curves represent the Earth-based (Sun-based) halo with the maximum allowed mass given in Eq.~\eqref{mstar}, and dashed contours denoting smaller halo masses as fractions of the Earth or the Sun mass. The black lines represent the reach of CASPEr-Electric assuming the standard background DM density, and the shaded regions are current constraints from astrophysics (green) and static EDM searches (gray). The left (right) panel represents Phase 1 (Phase 2) of the experiment, as defined by the input parameters given in~\cite{Budker:2013hfa}.}
	\label{fig:casper_e}
\end{figure}

In Figure~\ref{fig:casper_e}, we show a comparison between the reach of CASPEr, both for the axion halo and for the standard axion DM scenario. 
We use the parameters of Ref.~\cite{Budker:2013hfa}, which defines both Phase 1 and Phase 2 of the experiment. While the Phase 2 upgrade is necessary to probe a portion of the QCD axion band for the background DM model, Phase 1 would be sufficient to probe a 
range of QCD axion masses below $10^{-7}$ eV in the presence of an Earth-based of mass $M_\star \gtrsim 10^{-16}M_\oplus$, or a nearly-maximal Sun-based halo. The main reason for this is a higher energy density of a halo as well as the longer coherence time.

CASPEr is also able to probe the axial-vector coupling of axion to nucleon in a similar fashion, but without an external electric field (this version is known as CASPEr-Wind).
A spin-polarized material is similarly prepared, and an external magnetic field $\vec{B}_{\rm ext}$ is applied along the direction of the nuclear magnetic moment $\vec{\mu}_n$.
In the absence of the background axion field, there is no spin precession in the material because $\vec{\mu}_n \times \vec{B}_{\rm ext} = 0$. 
In the presence of the background axion, the material experiences additional torques, developing non-vanishing transverse magnetization. 
If the resonance condition is met, $2\mu_n B_{\rm ext} = m_\phi$, the amplitude of the induced transverse magnetization is resonantly enhanced as
\begin{equation}
M_\phi \simeq n\,p\,\mu_n\, |\vec{\mu}_n \times \vec{B}_\phi | t.
\end{equation}

We remind the reader that there are two contributions to the axion-induced magnetic field, $\vec{B}_\phi \propto \nabla \phi = \nabla_r \phi + \nabla_\perp \phi$, which are a radial and a tangential component of gradient (respectively).
As we have discussed in Sec.~\ref{sec:halo}, the radial component arises from the radial profile of axion halo, which is proportional to $\Rs^{-1}\simeq m_\phi v_\star$, while the tangential component arises from relative motion of experimental devices and axion halo, which is proportional to their relative velocity.

For the solar halo, the relevant relative velocity is given by Earth's orbital velocity around the Sun, roughly $v_{\rm rel}^{\odot} \simeq 10^{-4}$. The induced signal will be proportional to the torque $|\vec{\mu}_n \times \vec{B}_\phi|$, which depends on
\bea
|\vec{\mu}_n \times \vec{B}_\phi | \propto |\nabla \phi| \propto |\vec v_{\rm tot}| = |\vec v_\star + \vec{v}_{\rm rel}^{\,\odot}|.
\eea
Of course, the orientation of the nuclear spin with respect to the motion of the Earth depends on the time of day (as the Earth rotates) and the time of year (as the Earth moves through its not-quite-circular orbit around the Sun). For simplicity, in estimating sensitivity below we will take a fixed value $|\vec{v}_{\rm tot}| \sim v_\star + v_{\rm rel}^{\odot}$ (see Section \ref{app:modulation} for further details).

For the Earth-based halo, there  
is an analogous relative velocity originating in the rotation of the Earth through the axion halo.  
In the current instantiation of the CASPEr-Wind experiment, the nuclear magnetic moment $\vec{\mu}_n$ is oriented \emph{perpendicular} to the surface of Earth, implying that $\vec{\mu}_n\times\nabla\phi \simeq \vec{\mu}_n\times\nabla_\perp\phi$, and so the signal will come from the tangential gradient only. 
Therefore in our sensitivity estimates below, we use $\left|\nabla_\perp\phi\right| \simeq m_\phi\,\phi\,\left|\vec{v}_{\rm rel}^{\,\oplus}\right|$ with $\left|\vec{v}_{\rm rel}^{\,\oplus}\right| \simeq 1.5\times10^{-6}$. Note however that the ideal setup to optimize sensitivity to the axion halo would be to orient $\vec{\mu}_n$ parallel to the surface of Earth (maximizing the radial gradient) and perpendicular to Earth's lines of latitude (maximizing the tangential gradient also).
 
We determined the signal-to-noise ratio from the axion-spin coupling $g_{\phi NN}$ for both axion halo and background DM scenarios, and present the ratio $({\rm SNR})_{\rm halo}/({\rm SNR})_{\rm local}$ in Figure~\ref{fig:casper-w} rather than the absolute sensitivity; 
see Appendix~\ref{app:CE} for more details. For a solar halo, the 
ratio can be $1-2$ orders of magnitude even if $M_\star \lesssim 10^{-12}M_\odot$, whereas for the Earth halo it could exceed a factor of $10^5$.
For the projected sensitivities in the presence of local DM, we refer readers to~\cite{Graham:2013gfa,JacksonKimball:2017elr}.

\begin{figure}[t] \label{fig:casper-w}
\centering
\includegraphics[width=0.6\textwidth]{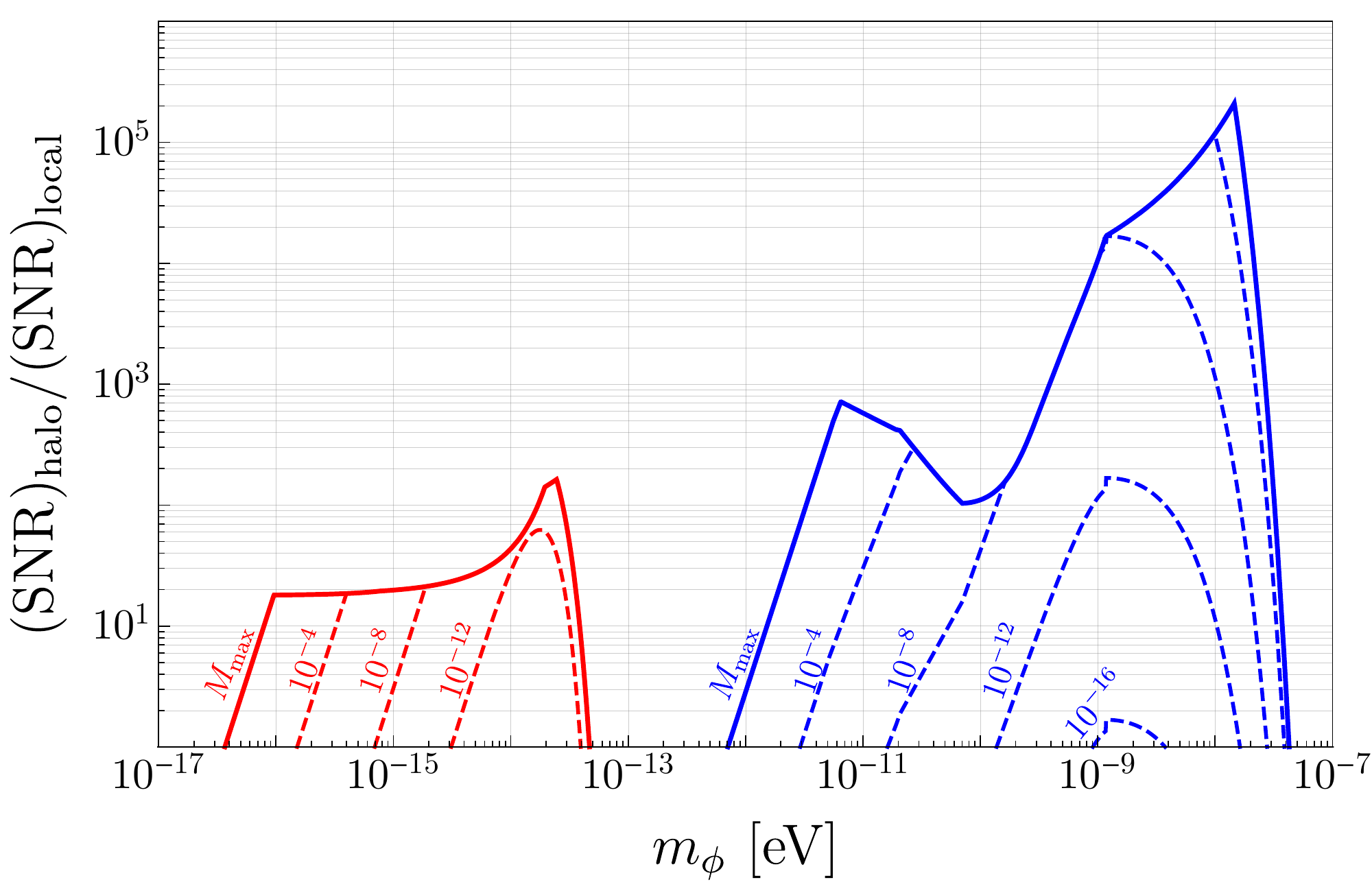}
\caption{ 
Signal-to-noise ratio for sensitivity to the coupling $g_{\phi NN}$, predicted in CASPEr-Wind in the presence of an axion halo, and shown relative to the case of background DM density. The thick blue (red) curve corresponds to a maximal axion halo bound to the Earth (Sun), with dashed contours denoting smaller halo masses as fractions of the Earth or the Sun mass. 
}
\end{figure}

\begin{figure}[t]
\centering
\includegraphics[width=0.49\textwidth]{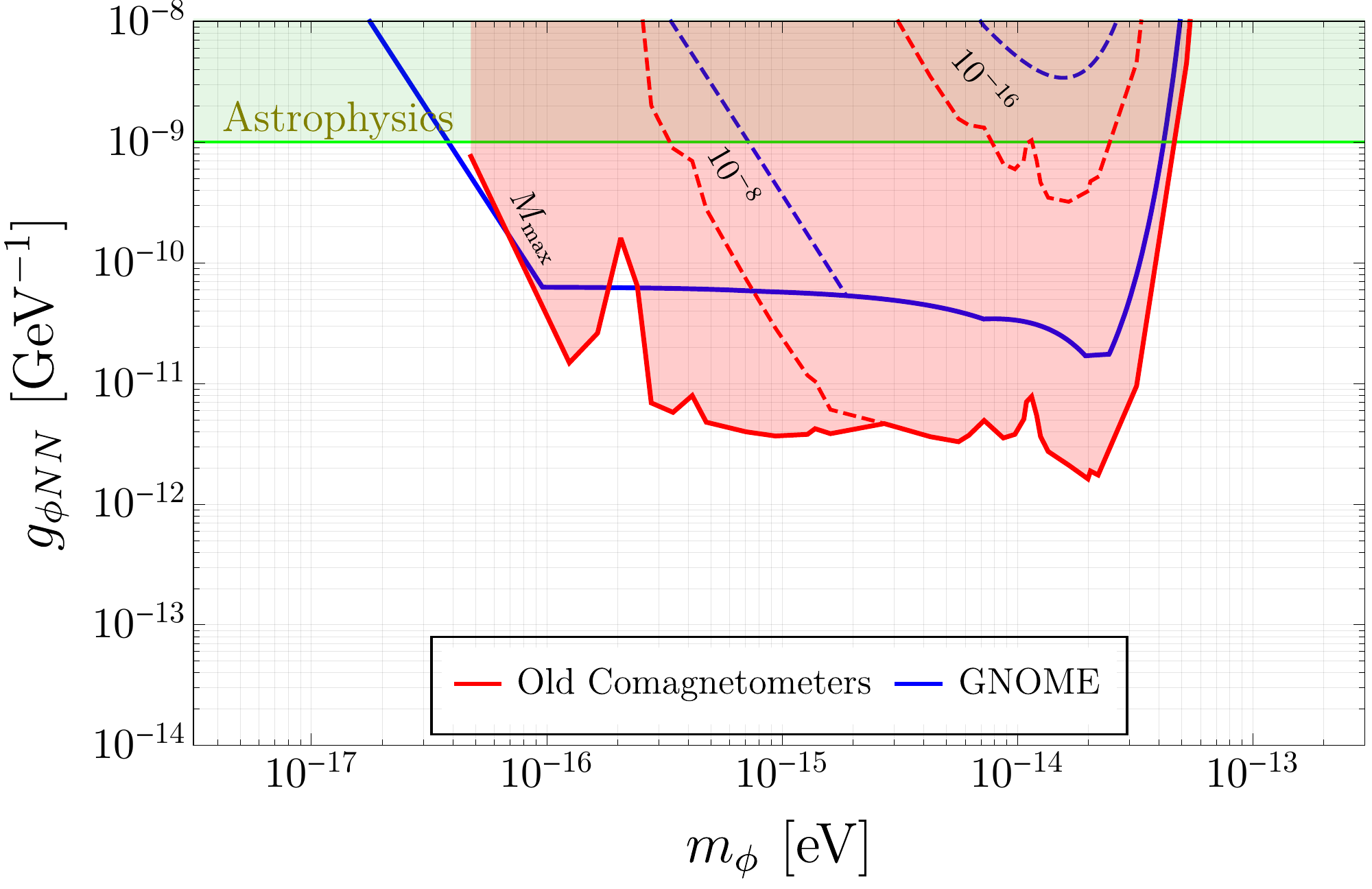}\,
\includegraphics[width=0.49\textwidth]{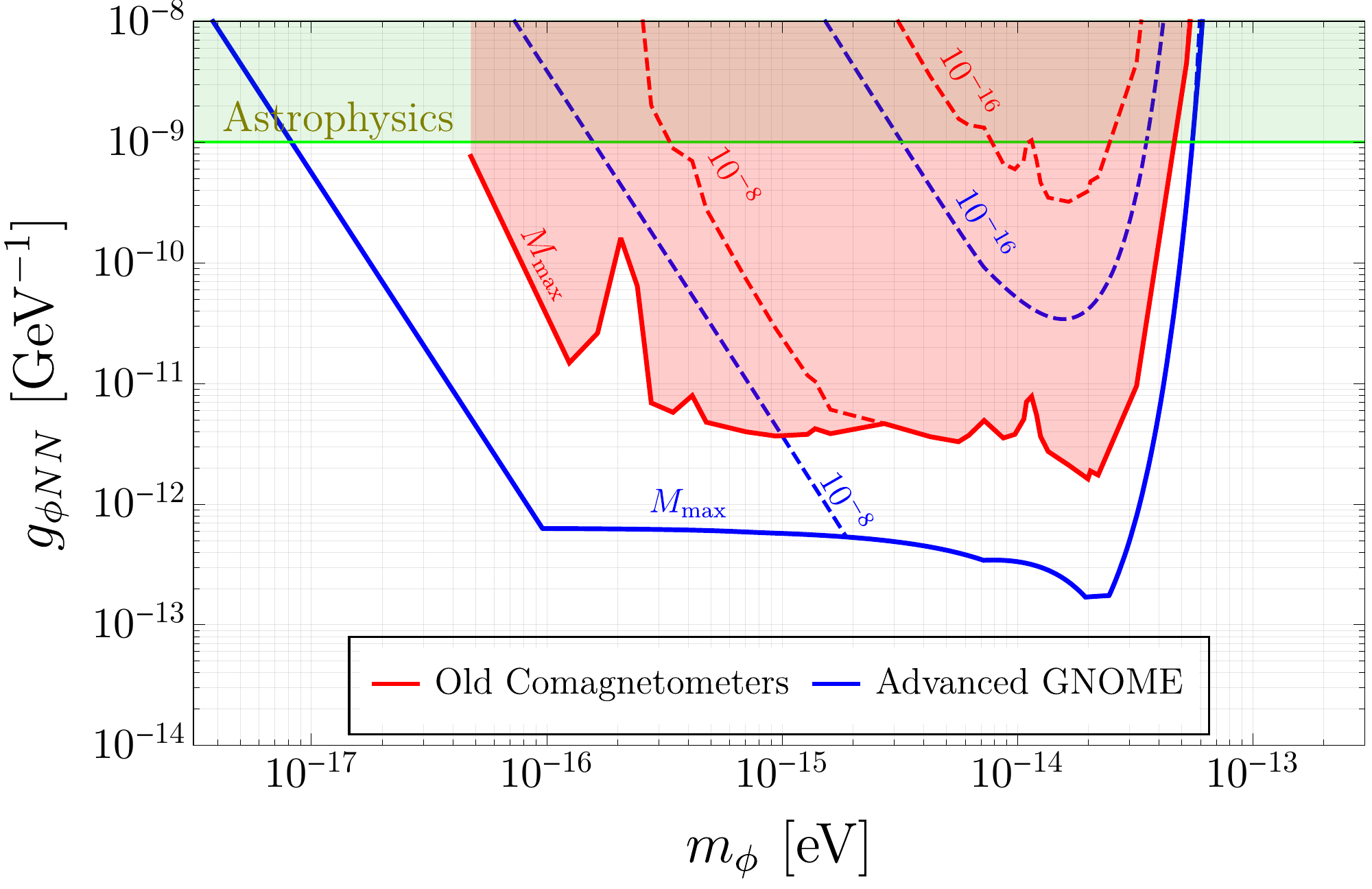}
\caption{Sensitivity of GNOME and ``old" comagnetometer experiments to the axion-nucleon spin coupling $g_{\phi NN}$. The blue solid line in the left (right) panel represents the current (Advanced) GNOME sensitivity, 
and the red shaded region refers to the constraint estimated in \cite{Bloch:2019lcy}, both interpreted in the presence of a maximal solar axion halo; the dashed contours of the same colours represent smaller halo masses as a fraction of a solar mass.
The green shaded region is the exclusion limit from astrophysical observations, specifically neutron star cooling and SN1987a, which are model-dependent, and both roughly at the level of $10^{-9}\,\GeV^{-1}$~\cite{Carenza:2019pxu}.}
\label{fig:comag_bound}
\end{figure}

\subsection{ALP Search using a global network of magnetometers}
We now consider experiments which are searching for ALPs using a set of magnetometers.
The Global Network of Optical Magnetometers to search for Exotic physics (GNOME) experiment consists of a network of magnetometers, and is optimized for searching transient signals. An example is the signal from encounters with axion stars with radius $10 R_{\rm \oplus} \lesssim R\lesssim 10^6 R_{\rm \oplus} $ and mass $M\ll 10^{-12}M_{\odot}\,$, such that the encounter rate with the Earth is $\Ocal(1/$year$)$ with a large overdensity~\cite{JacksonKimball:2017qgk}.\footnote{In the range of sensitivity of the GNOME experiment, the radius and the mass of axion star that allows encounter rate ${\cal O}(1/{\rm year})$ does not generally minimize the energy functional, indicating that it is not the stationary solution of the equation of motion.}
If the Earth encounters the axion stars with a relative velocity $v$, all GNOME sensors would register a transient signal within the time $T\sim 2 R_{\rm \oplus}/v$.

GNOME is sensitive to the axion coupling to nucleon spin~\cite{JacksonKimball:2017qgk}, which is given in Eq.~\eqref{CASPEr}.
In the nonrelativistic limit, the Hamiltonian can be written as, $H\simeq - \vec{\mu}_n \cdot \vec{B}_\phi \cos (m_\phi t)$, and thus a nuclear spin experiences an oscillating energy shift, $\Delta E = \mu_n B_\phi$ in the presence of an axion background. 
The signal-to-noise ratio for an optical magnetometer is
\bea
{\rm SNR} \simeq \frac{B_\phi}{S^{1/2}} F(t,\tau_\phi) 
\eea
where $S^{1/2}$ is the magnetometer sensitivity. 
The function $F(t,\tau_\phi)$ describes the behaviour of signal-to-noise as a function of total measurement time $t$ and a coherence time $\tau_\phi$, and it is given as
\bea
F(t,\tau_\phi) = 
\begin{cases}
\sqrt{t} & \textrm{for } t<\tau_\phi
\\
(\tau_\phi t)^{1/4} & \textrm{for } t>\tau_\phi
\end{cases}
\label{F}
\eea
where the behavior for $t>\tau_\phi$ is discussed in~\cite{Budker:2013hfa}. 
Given the current instantiation of GNOME, the magnetometer sensitivity is $S^{1/2}_{\rm current} =100 \, {\rm fT}/\sqrt{\rm Hz}\,$, whereas in the planned upgrade Advanced GNOME the sensitivity is expected to increase to $S^{1/2}_{\rm adv} = 1\, {\rm fT}/\sqrt{\rm Hz}\, $~\cite{JacksonKimball:2017qgk}. 

We only consider the solar axion halo since the GNOME experiment is sensitive to axion masses lighter than roughly $m_\phi \lesssim 10^{-13} \eV$.   
Below $m_\phi \lesssim 10^{-14} \eV$, the coherence time of the axion halo could be longer than a year (see Figure~\ref{fig:enhancement}), but we use the total measurement time to be a year for the estimate.  
Requiring SNR $\gtrsim1 $, we show the projected sensitivity of GNOME to $g_{\phi NN}$ in the presence of a solar halo in Figure~\ref{fig:comag_bound}.

We comment on three unique signatures that we may expect from GNOME, compared to other experiments we have considered. First, we have taken the sensitivity estimate of a \emph{single} GNOME (co)magnetometer, but the experiment consists in $N>10$ such detectors (with more coming online in the near future). Second, the axion halo that we are considering has macroscopic spatial coherence length, extending far beyond the radius of Earth's orbit, and thus, the signals in all GNOME stations would be coherent.
As a result, we expect that in the presence of a signal, the correlation between different GNOME stations will give an enhancement, likely improving the sensitivity proportionally to $\sqrt{N}$. Third, the multi-station design of GNOME lends itself perfectly to test a particular prediction of this scenario, namely, the existence of nearly-independent radial and tangential components of the gradient $\nabla\phi$. As of this writing, GNOME has stations around the world with different orientations relative to the surface of the Earth, allowing access to different proportions of the two gradient components. We emphasize that a dedicated search for a axion solar halo, including both the effect of detector latitude as well as orientation, would surely improve sensitivity to the signal as well.

In addition to GNOME, we also consider decade-old data from ${}^3$He-K noble-alkali comagnetometers \cite{Vasilakis:2011,Kornack:2005,Brown:2011}, which was recently used  to constrain the ALP coupling to nucleons $g_{\phi NN}$~\cite{Bloch:2019lcy}.
In the presence of an axion halo, the gradient measured by the ``old" comagnetometers is 
modified by the factor $\left(\nabla\phi / \nabla\phi_{\rm DM}\right) \simeq \left(\phi/\phi_{\rm DM}\right)\left(v_{\rm vir}/v_{\rm tot}\right)$, which are essentially fixed by the results of Figures \ref{fig:enhancement} and \ref{fig:gradenhancement}. 
In Figure~\ref{fig:comag_bound}, we show the constraint on $g_{\phi NN}$  
for the case of the solar halo (red lines). 
In the presence of a solar halo, the current GNOME sensitivity is weaker than these old data, except in a narrow region around $m_{\phi}\sim 2\times 10^{-16}\eV$ (see left panel); again, this may be improved in a dedicated analysis using the full power of all GNOME stations. 
With a planned upgrade, Advance GNOME will be sensitive over a wider range of masses $m_\phi$, and more sensitive to $g_{\phi NN}$ than the old comagnetometers by an order of magnitude or more. 
 
\subsection{ALP Search using an axion-induced current}

The third class of experiments we consider is based on the axion-induced effective current, which originates from the following anomalous coupling of axion to photon:
\bea \label{gagg}
\mathcal{L} \supset - \frac{1}{4} g_{\phi\gamma\gamma}\, \phi\ffdual,
\eea    
where $\widetilde{F}^{\mu\nu} = \eps^{\mu\nu\rho\sigma} F_{\rho\sigma}/2$ is a dual electromagnetic field strength. 
We treat $g_{\phi\gamma\gamma}$ as an independent parameter to analyse possible constraints. 
In the presence of the above anomalous coupling, one can write the Euler-Lagrange equation $\partial_\mu( F^{\mu\nu} + g_{\phi\gamma\gamma} \phi \widetilde{F}^{\mu\nu} ) = 0$ in terms of electromagnetic field as
\bea
\nabla\times \vec{B} = \frac{\partial\vec{E}}{\partial t} - g_{\phi\gamma\gamma} \left(\vec{E}\times \nabla\phi - \vec{B}\,\frac{\partial\phi}{\partial t} \right).
\eea 
From the above modified Maxwell's equation, it is clear that a background magnetic field $B_0$ would induce an effective current
\bea \label{jeff}
j_{\rm eff} =  g_{\phi\gamma\gamma} \sqrt{2\rho_{\star}} \cos\left(m_{\phi}t\right)B_0\,.
\eea
A Broadband/Resonant Approach to Cosmic Axion Detection with an Amplifying B-field Ring Apparatus (ABRACADABRA) has been proposed to utilize this axion-induced current to probe the axion DM~\cite{Kahn:2016aff}, and has recently set its first experimental limit on $g_{\phi\g\g}$ \cite{Ouellet:2018beu}. 
The experimental setup consists of a toroidal geometry generating a background magnetic field along the azimuthal direction, and a pickup coil coupled to a SQUID magnetometer, located at the center of a toroidal magnet.
The effective current $j_{\rm eff}$ is generated along $B_0$ according to Eq.~\eqref{jeff}, and due to Maxwell's equation, an oscillating component of magnetic field is developed orthogonal to $B_0$, whose flux is measured by a pickup circuit (for detailed experimental setup, see \cite{Kahn:2016aff}). 
The signal-to-noise is given as
\bea
{\rm SNR} \simeq \frac{|\Phi|}{S^{1/2}} F(t,\tau_\phi),
\eea
where the function $F$ is defined in Eq.~\eqref{F}, $\Phi$ is are the magnetic flux through SQUID, and $S^{1/2} = 10^{-6} \Phi_0 /\sqrt{\rm Hz}$ is the sensitivity of commercial SQUID with $\Phi_0 = h/(2e)$. 
The amplitude of the magnetic flux is approximately given as~\cite{Kahn:2016aff,Ouellet:2018beu}
\bea
\Phi \simeq g_{\phi\gamma\gamma} \sqrt{\rho_{\star}}\, B_{\rm max}\, V\, {\cal G}
\eea
where $V$ is the volume of the toroid, $B_{\rm max}$ is the maximum magnetic field inside the toroid, and ${\cal G}$ is a geometric factor. 

\begin{figure}[t]
	\centering
\includegraphics[width=0.49\textwidth]{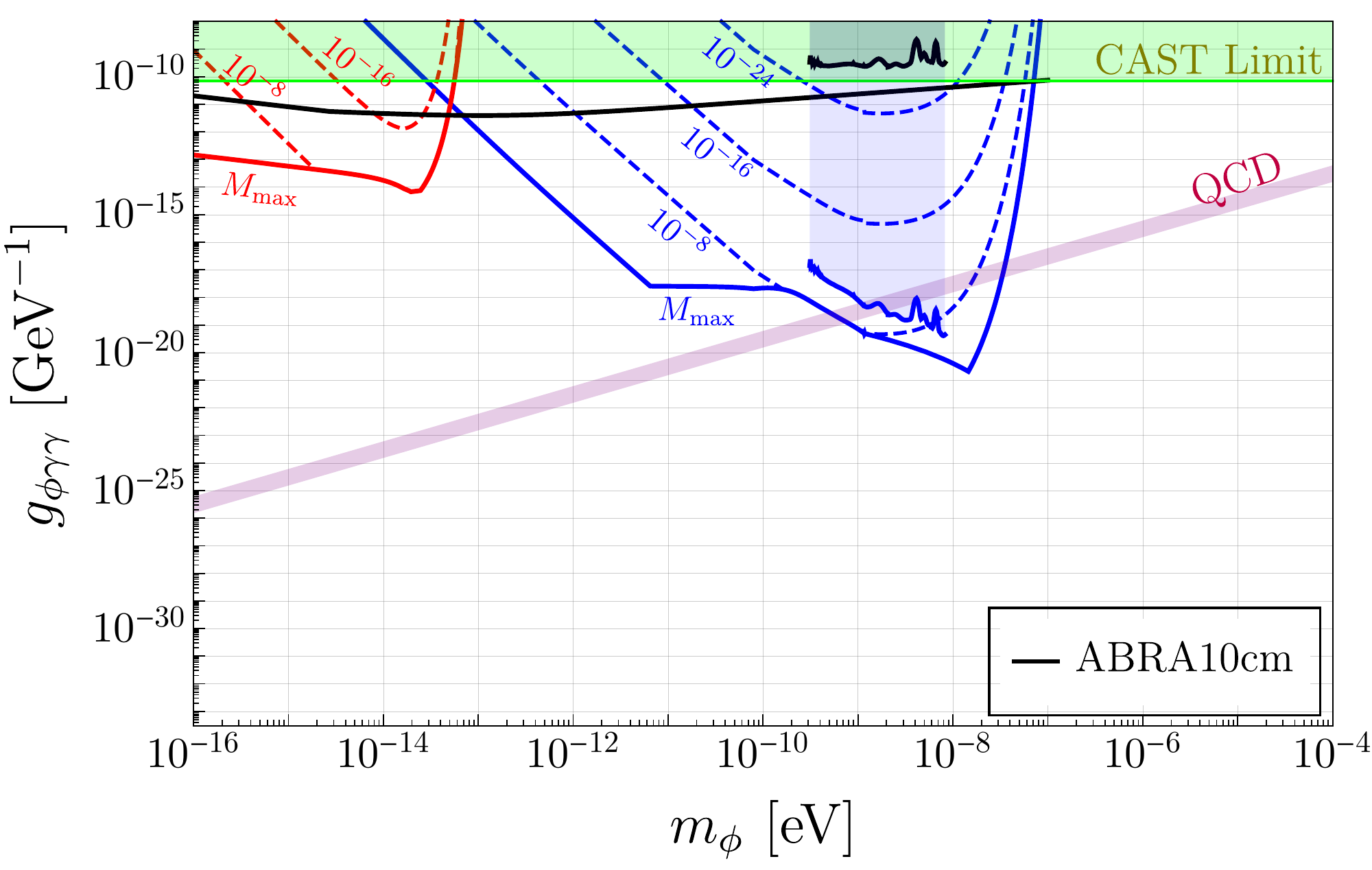}\,
\includegraphics[width=0.49\textwidth]{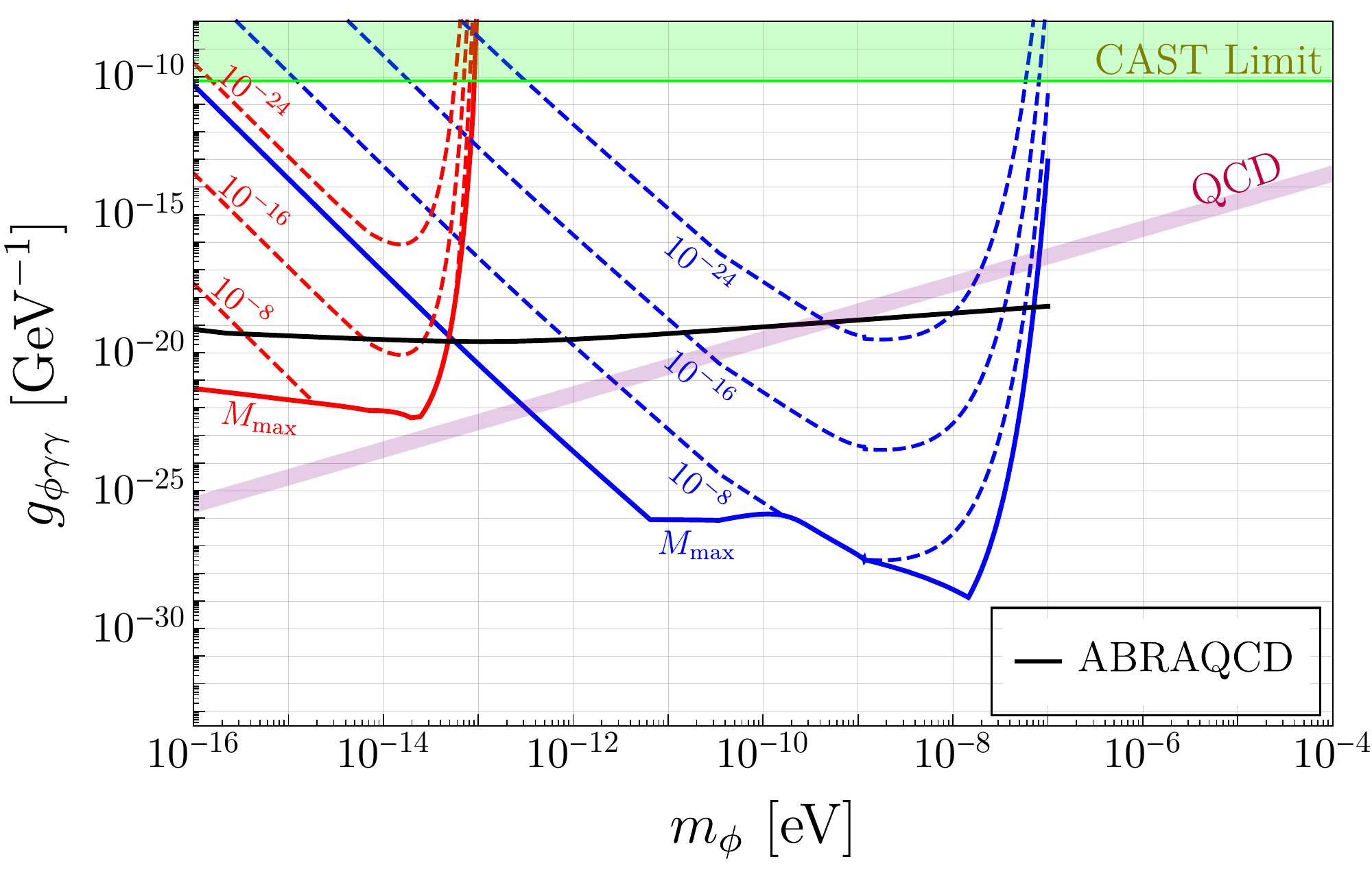}
	\caption{  
	Sensitivity to $g_{\phi\gamma\gamma}$ in the ABRACADABRA experiment.
	Black lines: projected sensitivity for background axion dark matter; thick blue lines: sensitivity for maximal Earth axion halo; thick red lines: sensitivity for maximal solar axion halo; dashed contours: sensitivity for axion halo masses smaller than maximal, labelled by the fraction of Earth or solar masses. The shaded regions represent the QCD axion band (purple), the current CAST constraint (green)~\cite{Anastassopoulos:2017ftl}, and the current ABRACADABRA constraint (black/blue). For ABRACADABRA, we take the same parameters used in~\cite{Kahn:2016aff} for a broadband search. In all cases we have also restricted the integration time to $1$ yr. 
		}
	\label{fig:broad_signals}
\end{figure}   

Using the parameters given in~\cite{Kahn:2016aff,Ouellet:2018beu} and requiring ${\rm SNR} \gtrsim 1$, we find current and projected sensitivities in Figure~\ref{fig:broad_signals}.
The solid lines correspond to the projected sensitivity on $g_{\phi \gamma\gamma}$ in the current (left panel) and optimistic eventual (right panel) iterations of the experiment; the latter implies a significant upgrade in toroid size, magnetic field strength, and integration time. 
The black lines assume the standard background DM density, whereas the blue (red) lines are for an Earth-based (Sun-based) axion halo, with dashed contours denoting masses smaller than the maximum given in Eq.~\eqref{mstar}. The black (blue) shaded region corresponds to the current constraint from ABRA10cm assuming local DM (maximal axion halo) parameters~\cite{Ouellet:2018beu}. It is intriguing that with the current level of sensitivity, ABRA10cm is already probing maximal axion Earth halos composed of QCD axions of mass $m_\phi \sim$ neV, which is several orders of magnitude better than existing limits from CAST~\cite{Anastassopoulos:2017ftl}.

The current and projected sensitivities shown in Figure~\ref{fig:broad_signals} are obtained for broadband searches.
Although ABRACADABRA may also conduct resonant searches, the current and near-future versions do not appear to be optimized for this mode of running, and thus, we have only shown the result for broadband search strategy.
It is worth noting that there exists another proposal, DM-Radio, searching for the same coupling constant $g_{\phi \gamma\gamma}$ based on the same principle. 
DM-Radio focuses on resonant search, claiming that this is the superior strategy on general theoretical grounds~\cite{Chaudhuri:2018rqn}. In any case, in the scenario of a bound axion halo the logic determining the relative benefits of broadband vs. resonant searches must be reconsidered carefully.
At present, there are no detailed estimates in publication regarding the DM-Radio sensitivity to light scalars, as the current focus of the collaboration is on the DM-Radio Pathfinder, a prototype experiment optimized to search for dark photons in a similar mass range \cite{Silva-Feaver:2016qhh}. For this reason, we do not show the projected sensitivity of DM-Radio in this paper, though the experiment can in principle enjoy the same large density and long coherence time in the axion halo scenario that we have described for ABRACADABRA.

\section{Constraining the Halo Mass} \label{sec:MaxMass}

In a predictive model, the ALP couplings to matter can be directly computed from the particle physics inputs; in that case, 
one can reinterpret the sensitivity of an experiment in terms of the maximum allowed axion halo mass $M_\star$. For example, the QCD axion, represented by the purple bands in the figures of the previous section, gives (up to model-dependent numerical prefactor) specific predictions for $g_{\phi\gamma\gamma}$, $g_d$, etc. at each choice of $m_\phi$. The experiments discussed above can thus constrain the maximum $M_\star$ allowed in particular halo scenarios.

If, for example, the ultimate sensitivity of ABRACADABRA (represented in the right panel of Figure \ref{fig:broad_signals}) sees no signal at an axion mass of $m_\phi = 10^{-8}$ eV, it would not only imply a constraint on the QCD axion as the background dark matter: it further rules out an Earth-based axion halo composed of such particles at the level of $M_\star \gtrsim 10^{-24} M_\oplus$.  This would represent a limit nearly $10$ orders of magnitude stronger than present-day gravitational measurements on DM overdensity around the Earth, derived from lunar laser ranging \cite{Adler:2008rq}.

\begin{figure}[t]
	\centering
	\hspace*{-0.9cm}
	\includegraphics[scale=0.55]{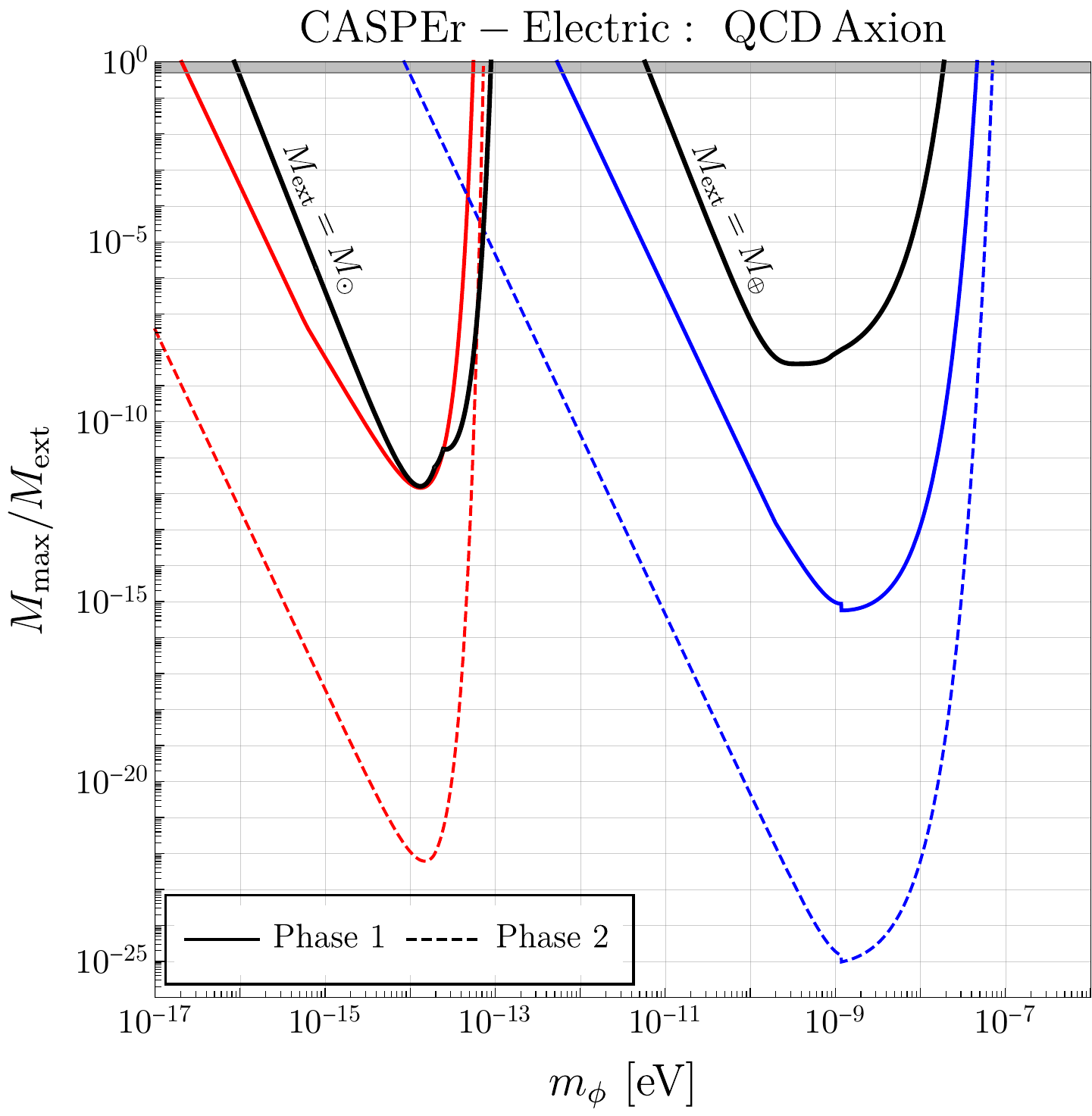}
	~
	\includegraphics[scale=0.395]{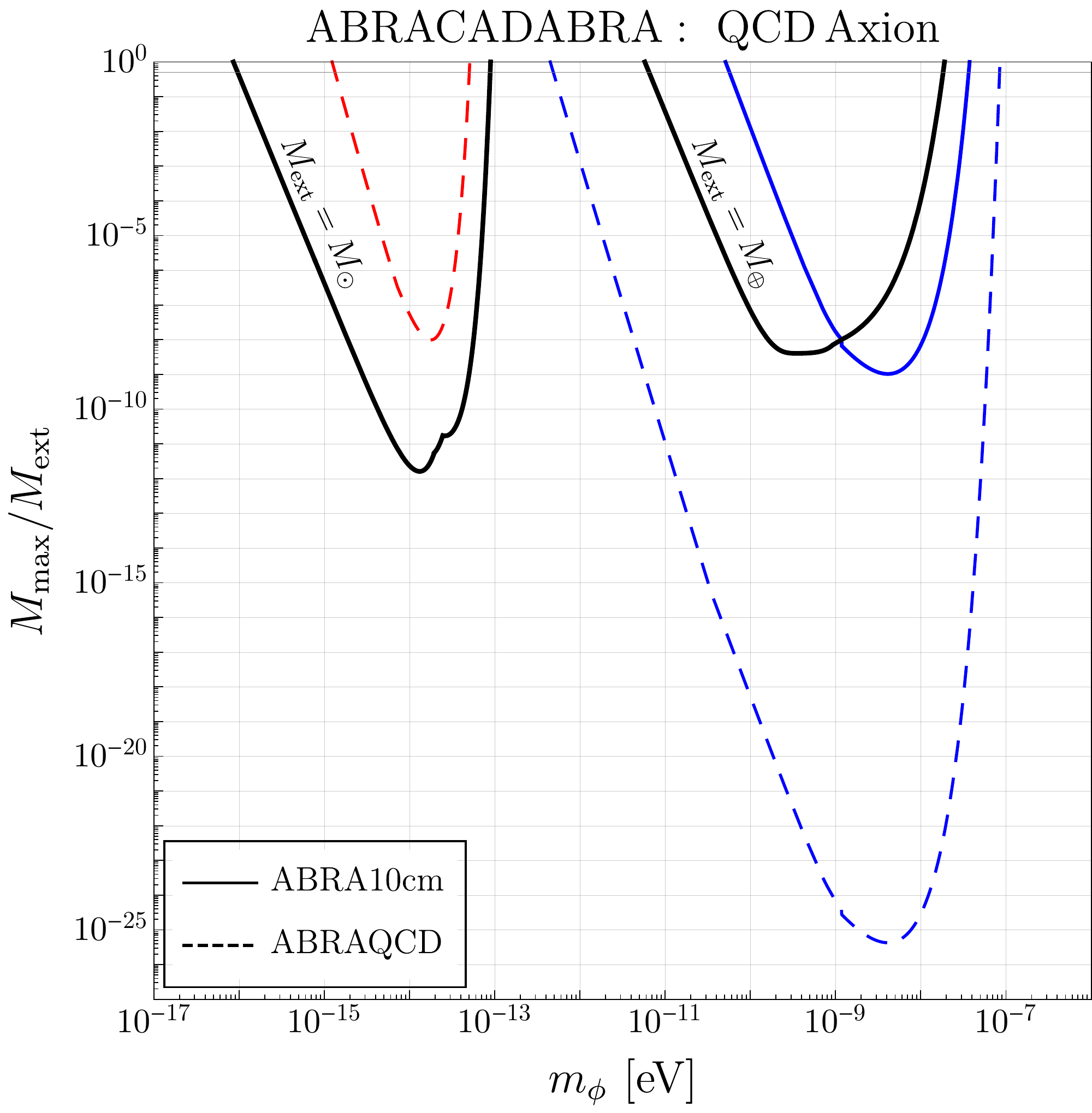}
	\caption{ Constraints on the maximum axion halo mass, assuming QCD axion couplings, for CASPEr-Electric measuring $g_d$ (left) and ABRACADABRA measuring $g_{\phi\gamma\gamma}$ (right). In both figures, the blue curves represent the Earth-based halo with $M_{\rm ext} = M_\oplus$, and the red curves are for the sun-based halo with $M_{\rm ext} = M_\odot$.
	}
	\label{fig:MMax}
\end{figure} 

More generally, we have recast the experimental sensitivities above as probes of QCD axion halos around the Sun and Earth as a function of $m_\phi$\footnote{Note that for $m_\phi \lesssim 10^{-10}$ eV, the QCD axion prediction is for a super-Plankian decay constant $f\gtrsim M_P$. We ignore the theoretical issues this may cause for the purposes of illustration in this work.}; the resulting constraints are given by the red and blue curves (respectively) in Figure \ref{fig:MMax}. 
Both CASPEr-Electric and ABRACADABRA, in their current iterations, are probing new regions of $M_\star$ with greater sensitivity than current gravitational measurements.  
For an Earth-based halo, future iterations of these experiments will reach sensitivity to halos of mass $M_\star \gtrsim 10^{-24} M_\oplus$;
furthermore, CASPEr-Electric can also constrain sun-based halos in Phase 2 of the experiment at the level of $10^{-22} M_\odot$.

\section{Outlook}\label{sec:conclusion}
We have analysed the effect of an axion halo, gravitationally bound to our solar system, on ALP DM searches based on pseudoscalar couplings to the SM particles. 
We have considered two cases, where the axion halo is hosted by the Sun and the Earth,
and we have shown that in each case current and near-future experiments searching for ALPs exhibit strong sensitivity to this scenario.
The sensitivity owes to the potentially large density of axion halo compared to local DM density, the modified gradients present in the axion halo, and an increased coherence time. The presence of each of these effects suggests that care should be taken in deciding experimental location, orientation, and (for resonant searches) shot times, as these are properties to which the axion halos we analysed are especially sensitive. These properties also represent new handles to distinguish background, virialized axion DM from axions which have become bound to the Earth or Sun.

The presence of an axion halo bound to the solar system can be distinguished from ordinary background DM by analysing directional information in searches for derivative couplings of the axions. When the signal depends on the gradient $\nabla\phi$, the orientation of the detector affects the amplitude of the expected signal, suggesting rotation of the apparatus as a non-trivial cross-check. The location (specifically, the latitude) of the experiment also has an effect, as the rotation of the Earth gives rise to additional gradients which are maximized closer to the Equator. Finally, for the solar halo, there are also important modulation effects on top of the usual classical oscillation of the field; the modulation on timescales of a solar day or a year represents yet another unique test of this scenario.

In a previous work~\cite{Banerjee:2019epw}, we performed an analogous analysis for fields in a similar mass range but with scalar couplings, those that give rise to effective oscillation of fundamental constants like the electron mass or fine-structure constant. Such oscillations are typically probed by atomic physics experiments, which are currently achieving impressive levels of sensitivity to low-amplitude oscillations. It is interesting to point out that some models of scalar field DM, notably the relaxion~\cite{Banerjee:2018xmn}, which is an axion-like particle that mixes with the Higgs due to the presence of CP violation~\cite{Flacke:2016szy}, predicts simultaneous presence of scalar and pseudoscalar couplings. It is possible that a combined search across different and diverse experimental programs may represent a unique probe of these models. 

A significant task to the scenario we have assumed in this work will be to outline a detailed formation history of scalar fields bound to objects in the solar system; thus far, we have only analysed the phenomenological consequences, should such a formation occur. On this point we merely mention that simulations have, only very recently, begun to shed light on the formation of the self-gravitating counterpart of our axion halos, known as boson stars. We speculate that, in the presence of a large external gravitational potential ({\it e.g.} that of the Sun), the very same relaxation processes that leads to boson star formation might instead lead to the formation of an axion halo (around the Sun), the latter being the ground state of the composite system. The details of this and other possible formation histories for bound axion halos will be pursued in the near future.

\section*{Acknowledgements}
We would like to thank Kfir Blum, Derek F. Jackson Kimball, Arran Phipps and Nicholas Rodd.
DB acknowledges the support by the DFG Reinhart Koselleck project, the European Research Council Dark-OsT advanced grant under project ID 695405, and the Simons and the Heising-Simons Foundations. His work was also supported in part by the DFG Project ID 390831469:  EXC 2118 (PRISMA+ Cluster of Excellence).
The work of JE is supported by the Zuckerman STEM Leadership Program.
The work of OM is supported by the Foreign Postdoctoral Fellowship Program of the Israel Academy of Sciences and Humanities.
The work of GP is supported by BSF, ERC, ISF, Minerva, the Segre award, and the Bessel Research award.

\appendix
\section{CASPEr} \label{app:CE}

CASPEr is a resonant search. 
At each frequency, we assume a measurement time as 
$t_s \approx \left(\frac{\Delta\omega}{m_\phi}\right) t_{\rm int}$, where $\Delta\omega$ is determined as $(\Delta \omega)^{-1} = 
{\rm Min} \left[ \tau_\phi , T_2\right]$. 
And, $t_{\rm int}$ is the total running time of the experiment, which is taken to be 3 years. 
For CASPEr-Electric, the maximal size of transverse magnetization is limited to
\bea \label{eq:MagCE}
M_\phi \simeq n\, p\, \mu_n\, d_a\, E \, \text{Min}[t_s,\tau_\phi,T_2].
\eea
The resonantly enhanced magnetization is measured by pickup look coupled to SQUID magnetometer. 
The magnetic flux through SQUID is estimated as
$
\Phi = (4\pi A_{\rm eff}) M_\phi
$\,,
where $A_{\rm eff} \simeq 0.3 \, {\rm cm}^2$ is effective geometrical area for the given experimental setup~\cite{Budker:2013hfa}.
On the other hand, the sensitivity (noise power spectral density) of commercial SQUID magnetometer is $S^{1/2} \simeq 10^{-6} \Phi_0 / \sqrt{\rm Hz}$ with $\Phi_0 = h/(2e)$. 
Therefore, the signal-to-noise ratio is
\bea
{\rm SNR} = \frac{\Phi}{S^{1/2}} F(t_s, \tau_\phi)
\eea
where $F$ is defined as \eqref{F}. 
Note that we have assumed that the sensitivity will not depend on the spin relaxation time $T_2$, because if $T_2 < \tau_{\phi}$ it is possible to recover coherent enhancement up to $t=\tau_\phi$ through the phase-cycling technique applied in the signal post-processing; see \cite{Garcon:2019inh} for details.
The projected sensitivities in the main text is obtained by requiring ${\rm SNR} \gtrsim 1$. 

The sensitivity of CASPEr-Wind can be obtained by the same procedure.
Since the noise sources are not explicitly discussed in~\cite{JacksonKimball:2017elr}, we instead compute the relative  
sensitivity in the halo scenario compared to local dark matter scenario for CASPEr-Wind.
The modification in sensitivities is estimated as
\bea \label{fratio}
\frac{(g_{\phi NN})_{\rm halo}}{(g_{\rm \phi NN})_{\rm local}}
=
\frac{\phi_{\rm DM}}{\phi}
\frac{v_{\rm vir}}{v_{\rm tot}}
\frac{[F(t_s,\tau_\phi) \text{Min}[t_s,\tau_\phi,T_2]]_{\rm local}}{[F(t_s,\tau_\phi) \text{Min}[t_s,\tau_\phi,T_2]]_{\rm halo}}.
\eea
We use the parameters of~\cite{Graham:2013gfa}, including $T_2 = 100$ sec, and the shot time ($t_s$) is chosen as discussed above for CASPEr-Electric. 

We emphasize that the axion mass being probed in CASPEr is proportional to the external magnetic field applied, so at low frequencies, it is necessary to use weak magnetic field to achieve the resonance effect discussed above. In this regime standard NMR techniques are not ideal, and so the sensitivity estimates we have used may not be appropriate. However, the apparatus  can be modified through the use of hyper-polarization techniques as well as modified encoding and detection methods not based on Faraday induction \cite{JacksonKimball:2017elr}. Such techniques are being employed in CASPEr-ZULF (Zero to Ultra-Low Field)~\cite{Garcon:2019inh,Wu:2019exd} as an extension of CASPEr-Wind, and will allow the experiment to probe axion masses in the range $10^{-22}$ eV $\lesssim m_\phi \lesssim 10^{-13}$ eV. In this case the phase-cycling technique mentioned above is crucial, since typically $\tau_\phi \gg T_2$ for the polarized materials used in the experiment. 

\bibliographystyle{h-physrev5}
\bibliography{ref}
\end{document}